\documentclass[floats,floatfix,showpacs,amssymb,prd,twocolumn,superscriptaddress,nofootinbib,longbibliography,reprint]{revtex4-1}

\usepackage{amssymb,amsmath,verbatim,mathtools,needspace,enumitem,etoolbox,graphicx,physics,microtype,afterpage,xspace,tabularx,lmodern,multirow}
\usepackage{gensymb}
\usepackage[normalem]{ulem}
\usepackage[dvipsnames, usenames]{xcolor}
\definecolor{linkcolor}{rgb}{0.0,0.3,0.5}
\usepackage[unicode, colorlinks=true, linkcolor=linkcolor, citecolor=linkcolor, filecolor=linkcolor, urlcolor=linkcolor, linktocpage, breaklinks]{hyperref}
\usepackage[all]{hypcap}
\usepackage[T1]{fontenc}
\usepackage[utf8]{inputenc}
\usepackage[usenames,dvipsnames]{xcolor}
\hypersetup{colorlinks=true,citecolor=romared,linkcolor=romared,urlcolor=romared}
\definecolor{romared}{RGB}{142,0,28}
\usepackage{aas_macros}
\usepackage{makecell}
\usepackage{soul}
\usepackage{booktabs}

\newcommand{\jhu}{\affiliation{William H. Miller III Department of Physics and Astronomy, Johns Hopkins University, 3400 North Charles
Street, Baltimore, Maryland 21218, USA}}

\begin{document}
	
\title{Gravitational wave inference of star cluster properties \texorpdfstring{\\}{} from intermediate-mass black hole mergers} 

\begin{abstract}
Next-generation ground-based gravitational wave observatories will observe mergers of intermediate-mass black holes (IMBHs) out to high redshift. Such IMBHs can form through runaway tidal encounters in the cores of dense stellar clusters. In this paper, we ask if the gravitational wave observation of a single merger event between two IMBHs, occurring in the aftermath of the coalescence of the clusters in which they formed, can be used to infer the properties of their host clusters, such as mass, redshift, and half-mass radius. We implement an astrophysically motivated analytic model for cluster evolution and IMBH growth, and we perform IMBH binary parameter estimation using a network of three next-generation detectors. We find that inferring the structural properties of clusters in this way is challenging due to model degeneracy. However, the posteriors on the cluster formation redshifts have relatively narrow peaks, and it may still be possible to infer the cluster formation history by measuring a whole population of IMBH binary merger events.
\end{abstract}

\author{Konstantinos Kritos}
\email{kkritos1@jhu.edu}
\jhu

\author{Luca Reali}
\jhu

\author{Ken K. Y. Ng}
\jhu

\author{Fabio Antonini}
\affiliation{Gravity Exploration Institute, School of Physics and Astronomy, Cardiff University, Cardiff, CF24 3AA, United Kingdom}

\author{Emanuele Berti}
\jhu

\date{\today}
\maketitle


\section{Introduction}
\label{sec:Introduction}

While the current LIGO-Virgo-KAGRA network of gravitational wave (GW) observatories routinely detects compact binary coalescences~\cite{LIGOScientific:2018mvr,LIGOScientific:2020ibl,KAGRA:2021vkt}, the community is developing next-generation (XG) ground-based GW observatories, such as the Einstein Telescope (ET)~\cite{Punturo:2010zz} and Cosmic Explorer (CE)~\cite{Evans:2021gyd}. With improved sensitivity at frequencies of the order of a few Hz, these detectors will observe the mergers of black holes (BHs) with masses in the hundreds of solar masses out to high redshift~\cite{Branchesi:2023mws,Gupta:2023lga,Fairhurst:2023beb,Reali:2024hqf}. As such, these detectors will give us access to the lower end of the intermediate-mass black hole (IMBH) regime, bridging the gap between stellar mass and supermassive BHs. Understanding the IMBH population and their environments is important, because IMBHs are believed to seed the growth of supermassive BHs~\cite{Volonteri:2012tp}.
 
There are several local optically selected IMBH candidates with masses $\sim 10^5$--$10^6M_\odot$ possibly lurking in the centers of dwarf galaxies~\cite{2020ARA&A..58..257G,Reines:2013pia,2019NatAs...3..755W,Mezcua:2017npy}. IMBHs may also power some ultraluminous X-ray sources, such as M82X-1 in the starburst galaxy M82, which has been interpreted as an accreting IMBH of $\simeq400M_\odot$~\cite{2014Natur.513...74P} (but see~\cite{Brightman:2016pax} for an alternative interpretation of this source as a $\sim26M_\odot$ BH accreting beyond the Eddington limit). 
The hyperluminous X-ray source ESO243-49 HLX-1, with an inferred BH mass of at least $500M_\odot$~\cite{2009Natur.460...73F} and its association with a massive ($\sim10^6M_\odot$) young star cluster~\cite{2012ApJ...747L..13F}, is arguably the strongest IMBH candidate.
Current radio observations of Galactic globular clusters set an upper limit on the mass of putative IMBHs accreting at the Bondi rate in their centers to $\lesssim1000M_\odot$~\cite{Tremou:2018rvq}, although kinetic data support the presence of IMBHs in some systems~\cite{Gebhardt:2005cy,2024arXiv240506015H,Huang:2024gpv}. 
Unlike supermassive BHs, confident detections of IMBHs based on stellar dynamics are still elusive due to their smaller radius of influence, which scales proportionally to BH mass. 
Ultimately, the unequivocal detection of an IMBH requires a BH mass measurement in the range $100$--$10^6M_\odot$. 
Given the current challenges with dynamical and electromagnetic probes, a promising alternative avenue to measure IMBH masses is the observation of GWs from the inspiral and merger of a BH binaries. 
The analysis of gravitational waveforms provides a clean method to infer IMBH masses and spins, not plagued by the uncertainties related to accretion physics or stellar dynamics. 

There is a multitude of proposed IMBH formation scenarios, that generally fall into one of the following three categories (see Fig.~1 in Ref.~\cite{2020ARA&A..58..257G}): (i) direct collapse of low-metallicity gas clouds, (ii) growth of BH remnants of metal-poor Population III (Pop. III) stars through gas accretion,
and (iii)  growth of an initial seed through mergers. The latter scenario includes three types of mergers in dense star clusters: repeated BH mergers \citep{Antonini:2018auk}, repeated stellar mergers \citep{PortegiesZwart:2004ggg}, or repeated accretion of stars by a BH, i.e., tidal disruption events (TDEs)~\cite{Rizzuto:2022fdp}.
An important difference between these scenarios is that channels (i) and (ii) are believed to operate only at high redshift ($z\gtrsim 10$), when heavy elements are not so abundant, as they are only effective at low metallicities. 
On the other hand,  the third scenario may occur at any redshift, as long as sufficiently compact clusters form throughout cosmic time.
Therefore, mergers of binaries with IMBH components formed via channels (i) and (ii) may be dominant in the high-redshift Universe, while channel (iii) might produce merger events even at lower redshifts.

Several mechanisms can lead to the formation and merger of IMBH-IMBH binaries.
One of the scenarios for BH growth is the hierarchical assembly of BHs, first from the stellar mass to the intermediate mass regime, and then to the supermassive regime~\cite{Ebisuzaki:2001qm,Sedda:2019btz,2021MNRAS.502.2682A}. 
In this scenario, IMBHs first form from gravitational runaways 
in the cores of dense star clusters.
Clusters that reside within a few kpcs from the center of their host galaxy 
undergo dynamical friction and can sink into the core before they ``evaporate'' \citep{Antonini:2012sj}, where they merge with the nuclear star cluster or with other inspiralling globular clusters (see, e.g., Sec.~7.2 of Ref.~\cite{2020A&ARv..28....4N}). 
Dynamical friction also contributes to the growth of the central nuclear star cluster~\cite{Antonini:2012sj,Gnedin:2013cda}. 
In this way, IMBHs assembled in different cluster environments are brought together as their host clusters carry them along, eventually ending up in the central regions of the galaxy~\cite{2018MNRAS.477.4423A}.
For example, the compact stellar complex IRS 13E is hypothesized to be the core of a dissolved young star cluster hosting an IMBH that sank into the Galactic center.
However, the absence of a radio or x-ray signature casts doubt on the presence of an IMBH at the center of this system~\cite{2005ApJ...625L.111S}, that could be dominated by a subcluster of stellar-mass BHs~\cite{Banerjee:2011fa}.
A second possible scenario leading to the formation of IMBH-IMBH pairs is the merger of ultradwarf galaxies.
Dynamical friction brings the IMBHs into the central regions, and the sinking timescale can be significantly shorter if the IMBHs are embedded in NSCs~\cite{Palmese:2020xmk}.
According to observations in the local Universe, the NSC occupation fraction (i.e., the fraction of systems with identified NSCs)
is at least 80\% in the galaxy stellar mass range $\sim10^8$--$10^{10}M_\odot$: see the right panel of Fig.~3 in~\cite{2020A&ARv..28....4N,2024arXiv240310602R}.
In a third scenario, star clusters can form in the clumpy star-forming turbulent environment of giant molecular cloud complexes, and then merge hierarchically after sinking into the central regions. This process can be very efficient and rapid, as suggested by magnetohydrodynamic simulations~\cite{Shi:2024wgh}. 
In particular, IMBHs can be brought together and form a binary much more efficiently when surrounded by stellar clusters, as the dynamical friction timescale can be effectively reduced by orders of magnitude for the cluster-IMBH system due to its higher dynamical mass (see also Fig.~3 of~\cite{Palmese:2020xmk}). 
The simulations of Ref.~\cite{Khan:2021jqf} suggest that the formation and hardening of IMBH-IMBH pairs in the centers of dwarfs is very efficient and leads to their merger. 

There are two broad classes of gravitational runaway channels that can form IMBHs in star clusters, based on the timescale over which they operate: the ``fast'' and ``slow'' mechanisms~\cite{2015MNRAS.454.3150G}.

The fast channel operates early in the cluster's evolution, and corresponds to either repeated stellar collisions or successive BH-BH mergers.
Stellar collisions require very dense collisional systems with a small enough initial relaxation time to allow for core-collapse before massive stars evolve to BHs.
They result in the formation of a massive star (which then collapses into an IMBH) within $\sim10\,\rm Myr$~\cite{Gurkan:2005xz,2018MNRAS.478.2461G,Kritos:2022non}. 
For heavier stars the collisional runaway scenario is likely to be limited by strong stellar winds, which could lead to a mass loss rate that exceeds the growth rate, especially in high-metallicity systems~\cite{Mapelli:2016vca,2009A&A...497..255G}. 
Therefore, this channel may only be relevant in low-metallicity environments, because metal-poor stars are expected to have a weaker wind-driven mass loss.
The second possible route for the fast channel (i.e., successive BH-BH mergers) is limited by the escape velocity of the cluster. The relativistic kick imparted to the merger remnant could reach hundreds of $\rm km\, s^{-1}$, leading to an ejection in all but the heaviest clusters (such as the progenitors of ultracompact dwarfs, or nuclear star clusters), that have a high enough escape velocity ~\cite{Holley-Bockelmann:2007hmm}.

In the slow channel, after most BHs have been ejected from the system, the subcluster of the few remaining BHs becomes Spitzer-stable, and couples efficiently to the rest of the cluster. A runaway BH then begins to grow through the consumption of low-mass stars.
It turns out that the mass of the growing BH asymptotes to a value that depends on the properties of the host environment~\cite{Stone:2016ryd,Rizzuto:2022fdp}.
Such a growing BH from runaway tidal encounters is unlikely to be ejected, because relativistic kicks do not operate, and (by momentum conservation) BH-star interactions impart an insignificant Newtonian recoil to the BH (see Sec.~4.3 of Ref.~\cite{Rizzuto:2022fdp}).

Assuming that binaries of IMBHs build up from the coalescence of inspiralling globular clusters in the centers of galaxies, a precise characterization of the IMBH masses and spins may allow us to connect them with the properties of their cluster progenitors.
Recent studies have shown that XG detectors can measure the properties of IMBHs out to a high redshift $z$~\cite{Gupta:2023lga,Fairhurst:2023beb,Reali:2024hqf}. In particular, binaries with IMBH component masses $\lesssim 1000M_\odot$  at cosmic noon ($z=2$) produce GWs with signal-to-noise ratio (SNR) in the hundreds, such that their masses and redshift can be measured with percent-level accuracy.
There are also proposals to constrain the properties of star clusters using GW measurements of stellar-mass BH merger events with current ground-based detectors; however, these inferences are prone to contamination from the isolated binary evolution channel, as well as other channels operating in the stellar-mass regime~\cite{Romero-Shaw:2020siz,Ng:2023wbx,Fishbach:2023xws}.

\begin{figure}
    \centering
    \includegraphics[width=0.5\textwidth]{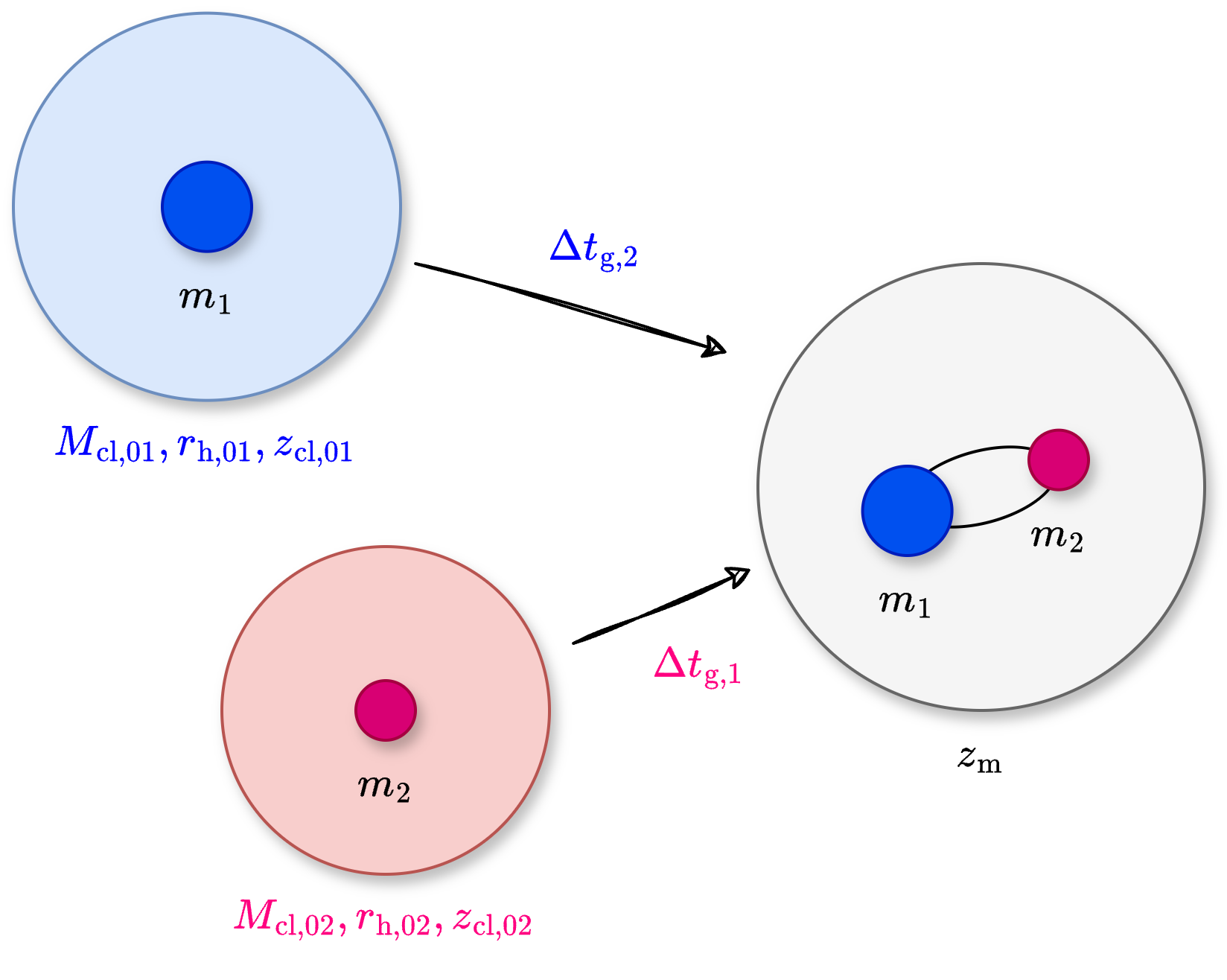}
    \caption{A sketch of the scenario examined in this paper: the formation of IMBH binaries with mass components $m_1$ and $m_2$ and their subsequent coalescence at redshift $z_{\rm m}$ after the merger of their host star clusters. Each star cluster has an initial mass $M_{\rm cl,0i}$, initial half-mass radius $r_{\rm h,0i}$, and formation redshift $z_{\rm cl,0i}$, where $i=1,2$. The symbol $\Delta t_{\rm g,i}$ denotes the time between the formation of each cluster and their merger, while $\Delta t_{\rm d}$ will denote the delay time between the merger of the clusters and the merger of the IMBH binary.}
    \label{Fig:parameter_description}
\end{figure}

In this paper, we assess how the detection and parameter estimation (PE) of individual IMBH-IMBH mergers in XG observatories can be exploited to infer the properties of their host stellar clusters. We consider several different binary IMBHs with total (source-frame) mass $\lesssim1500 M_\odot$ and redshift $\lesssim 8$. For each of these systems, we generate the corresponding GW signal with state-of-the-art waveform models and perform full Bayesian PE in a network of XG detectors. We develop an astrophysically motivated analytical model that relates the mass of the IMBH to the initial conditions of the evolving cluster as a function of time, assuming that the BH grows through successive TDEs.
Clusters in the tidal environment of a galaxy, or in a giant star-forming complex,
inspiral by dynamical friction and deposit their formed IMBHs into the central regions, where they pair up with another IMBH that formed in another cluster that also sank into the center. 
In alternative, IMBHs assembled in the centers of dwarf galaxies can be brought together and form a bound pair after their parent galaxies have merged (see Fig.~\ref{Fig:parameter_description} for a schematic representation).
For each IMBH merger, we then combine the PE results and our analytical model to hierarchically infer the mass, half-mass radius, and formation redshift of the progenitor host clusters.

The paper is structured as follows.
In Sec.~\ref{sec:Black_hole_growth_model} we develop our astrophysical model for BH growth and cluster evolution. 
In Sec.~\ref{sec:Single-event_cluster_posteriors} we present the cluster posterior distributions for a set of simulated IMBH binary events.
In Sec.~\ref{sec:Limitations_of_the_model} we discuss the limitations of our model. 
Finally, in Sec.~\ref{sec:Conclusions} we present our conclusions and directions for future work.
To improve readability, some technical material is presented in Appendix. In Appendix~\ref{App:Tidal_capture_radius} we discuss our calculation of the tidal capture radius;
in Appendix~\ref{App:Cluster_dissolution_and_BH_feedback} we compute the conditions for cluster dissolution and the effect of BH feedback; in Appendix~\ref{App:Parameter_estimation_of_IMBH-IMBH_binaries} we briefly discuss our parameter estimation calculation; in Appendix~\ref{App:mass_spin_evolution} we demonstrate that the spin of a BH growing through runaway stellar consumption asymptotes to zero; and in Appendix~\ref{App:Margninalized_cluster_posteriors_for_fs003} we present marginalized posteriors of the cluster properties in some specific cases.

\section{Black hole growth model}
\label{sec:Black_hole_growth_model}

A model for the evolution of the BH mass $M_{\rm BH}$ as a function of time $t$ given the cluster parameters is required to relate the mass of the growing BH to the properties of its host cluster.
Since the cluster expands and loses mass as it evolves, we define this dependence through the cluster's global initial conditions: the initial cluster mass $M_{\rm cl,0}$, the initial half-mass radius $r_{\rm h,0}$, and the redshift of cluster formation $z_{\rm cl,0}$.
In this section we present the central BH growth rate formula (Sec.~\ref{Sec:Stellar_conpsution_rate}), the evolution of the cluster's global properties (Sec.~\ref{Sec:Star_cluster_evolution}), an analytic time-dependent BH mass solution (Sec.~\ref{Sec:Black_hole_mass_growth}), an investigation of how the BH growth is affected by the cluster's initial conditions (Sec.~\ref{sec:Dependence_on_cluster's_initial_condition}), and a calculation of the delay time $\Delta t_{\rm d}$ between cluster coalescence and IMBH binary merger (Sec.~\ref{Sec:Delay_time}). 

\subsection{Stellar consumption rate}
\label{Sec:Stellar_conpsution_rate}

Assuming most of the cluster's mass is locked in light stars that survive for billions of years, the stellar number is $N_{\star}\simeq M_{\rm cl}/\overline{m}_\star$ where $\overline{m}_\star\simeq0.6M_\odot$ for a Kroupa initial mass function~\cite{Kroupa:2002ky}.
We neglect the contribution of BHs (and of the central BH, of mass $M_{\rm BH}$) 
to the total mass of the cluster, as the former can be shown to contribute no more than $\simeq10\%$ of the total mass~\cite{Kritos:2020fjw}.
According to the virial theorem, the root-mean-square velocity in the cluster is $v_{\rm rms}\simeq(0.4GN_\star \overline{m}_\star/r_{\rm h})^{1/2}$~[see, e.g., Eq.~(2) from Ref.~\cite{1971ApJ...164..399S}]. 
We further assume isothermal and isotropic conditions within the half-mass radius, so that the velocity dispersion in the cluster core is $\sigma=v_{\rm rms}/\sqrt{3}$.
The influence radius of the central BH is defined as $r_{\rm a}\equiv GM_{\rm BH}/\sigma^2$, and it corresponds approximately to the radius within which the potential of the BH starts to dominate over the stellar component. It also roughly corresponds to the distance from the center which encloses a total stellar mass of $\approx M_{\rm BH}$.
Stars within $r_{\rm a}$ can have bound orbits around the central BH.

When a star gets close to the central BH, the tidal forces can be strong enough to rip the star apart in a TDE.
The critical radius for a TDE to occur (also known as the ``tidal radius'') is given by $r_{\rm T}\simeq2\times10^{-6}{\,\rm pc}~[M_{\rm BH}/(100M_\odot)]^{1/3}$ for main-sequence solar stars: see Eq.~(6.2b) from Ref.~\cite{Merritt2013} with $\eta=0.844$ for main sequence stars, where we used the solar value for the radius of the star.
Stars on orbits with a pericenter larger than $r_{\rm T}$ can still be captured around the BH on more tightly bound orbits through the deposition of orbital energy into (tidally excited) stellar internal modes~\cite{1975MNRAS.172P..15F,1977ApJ...213..183P}.
We estimate the capture radius to be $r_{\rm C}=\beta r_{\rm T}$ with $\beta=2.0$ (see Appendix~\ref{App:Tidal_capture_radius} for details).
The tidal dissipation experienced at every pericenter passage causes the orbit to decay and circularize, until the star is eventually disrupted when it reaches $r_{\rm T}$.

Reference~\cite{Rizzuto:2022fdp} showed that the tidal inspiral of a captured solar-mass star is an efficient process, rapidly leading to tidal disruption within $<100\,\rm yr$ for $M_{\rm BH}=50M_\odot$.
Therefore, we assume that stars that encounter the central BH with a pericenter $\le r_{\rm C}$ are consumed and lost from the core of the cluster.
Stellar trajectories with a pericenter $\le r_{\rm C}$ are called ``loss-cone orbits'' because the space of velocity vectors tangent to these orbits defines a cone structure [see, e.g., Fig.~6.2(a) of Ref.~\cite{Merritt2013}].
However, not all stars on loss-cone orbits are lost if their velocity vector diffuses out of the loss-cone on a timescale of less than one orbital period.
In general, there is a critical distance from the central BH within which two-body relaxation processes cannot perturb orbits outside the loss-cone within an orbital time~\cite{1976MNRAS.176..633F}.
We have checked that for the parameter space of interest, most of the flux comes indeed from within the influence radius $r_{\rm a}$ (see the lower-panel of Fig.~6.5 from~\cite{Merritt2013}). 
Thus, for systems containing an IMBH, most of the BH's stellar flux comes from within this critical radius.

We follow Ref.~\cite{Rizzuto:2022fdp} and use $2r_{\rm T}$ for the loss-cone radius. Suppose tidally captured stars strongly bound to the BH do not contribute to the growth of the BH. In that case, we are overestimating the consumption rate by a factor of two (because the cross section depends linearly on the loss-cone radius in the gravitational focusing regime).

An estimate for the rate of stellar consumption by the central BH in the cluster is given in Refs.~\cite{1976MNRAS.176..633F,1977ApJ...211..244L} in the context of the full loss-cone model, which assumes instantaneous repopulation of loss-cone orbits. However, the full loss-cone model overestimates the consumption rate: the replenishment of stellar orbits extremely close to the BH does not occur immediately, but it normally requires more than an orbital time. Furthermore, those close orbits release a lot of energy into the cluster, leading to a rapid expansion of the core, which causes a feedback effect that limits the growth of the BH~\cite{1977ApJ...217..281S}.
A more conservative estimate is given in Eq.~(6.15) from Ref.~\cite{Merritt2013}, where the assumption is that all loss-cone orbits within $r_{\rm a}$ are replenished within the orbital time at that radius.
The equilibrium star density follows a cuspy distribution near the BH with a profile $n_\star\propto r^{-\gamma}$. A zero energy flow solution implies the Bahcall-Wolf law with $\gamma=7/4$~\cite{1976ApJ...209..214B}, as verified in Ref.~\cite{Preto:2004kd} with direct $N$-body simulations (see also Fig.~6 of Ref.~\cite{Rizzuto:2022fdp}).
With this choice, the consumption rate is given by
\begin{subequations}
\begin{align}
    \Gamma_{\rm C}&\simeq7{r_{\rm C}\over r_{\rm a}}{M_{\rm BH}\over \overline{m}_\star}\sqrt{{GM_{\rm BH}\over r_{\rm a}^3}}\\&\simeq 53\,{\rm Myr}^{-1}\left({M_{\rm BH}\over100M_\odot}\right)^{-2/3} \left({\sigma\over{10\,\rm km\, s^{-1}}}\right)^{5}\label{Eq:consumption_rate_b}
\end{align}%
\label{Eq:consumption_rate}%
\end{subequations}%
for solar mass stars.

\subsection{Star cluster evolution}
\label{Sec:Star_cluster_evolution}

According to H\'enon's principle, the rate of energy production (and hence the stellar consumption rate) is regulated by the efficiency of two-body relaxation processes.
The half-mass relaxation time is given by $\tau_{\rm rh}\simeq67\,{\rm Myr}\,(N_\star/10^5)^{1/2}(r_{\rm h}/{\rm pc})^{3/2}$: see Eq.~(10) from~\cite{Antonini:2019ulv}, with $\ln\Lambda=10$ and $\psi=1$.
A fraction $\zeta\simeq0.0926$ of the cluster's energy can be transferred throughout within $\tau_{\rm rh}$, such that $\dot{E}_{\rm cl}=-\zeta E_{\rm cl}/\tau_{\rm rh}$~\cite{Gieles:2011wh}.
The rate of energy generation is independent of the microphysical details of the heat production mechanism and commences right after core collapse, which occurs at time $\tau_{\rm cc}\simeq3.21\tau_{\rm rh}$~\cite{Antonini:2019ulv}. 
Furthermore, the cluster expands as a consequence of adiabatic mass loss due to stellar evolution.
We assume that the mean stellar mass $\overline{m}_\star$ evolves with time according to $d\overline{m}_\star/dt = -\nu \overline{m}_\star / t$ for $t>\tau_{\rm se}=2\,\rm Myr$, with $\nu=0.07$~\cite{2014MNRAS.442.1265A,Antonini:2019ulv}.
Combining the balanced evolution condition with the time derivative of the relation $E_{\rm cl}\simeq-0.5M_{\rm cl}v_{\rm rms}^2$ for the cluster energy results in an equation for the evolution of the half-mass radius (compare Eq.~(15) from~\cite{Antonini:2019ulv}):
\begin{align}
    {dr_{\rm h}\over dt}=\zeta {r_{\rm h}\over\tau_{\rm rh}}\Theta(t-\tau_{\rm cc}) + 2{r_{\rm h}\over M_{\rm cl}}{dM_{\rm cl}\over dt} + {\nu r_{\rm h}\over t}\Theta(t - \tau_{\rm se})\,.
    \label{Eq:radius_evolution_equation}
\end{align}
The last term considers the effect of stellar mass evolution, and $\Theta$ is the Heaviside function.
With our approximation $M_{\rm cl}\simeq \overline{m}_\star N_\star$, we have $dM_{\rm cl}/dt\simeq \overline{m}_\star dN_\star/dt + N_\star d\overline{m}_\star /dt$.
Since the velocity distribution thermalizes and the high-velocity tail is replenished roughly every $\tau_{\rm rh}$, the cluster slowly evaporates as it expands~\cite{1985IAUS..113..521A}.
We denote by $\xi_{\rm e}$ the fraction of ejected stars with a velocity larger than the escape velocity, such that $dN_\star/dt=-\xi_{\rm e}N_\star/\tau_{\rm rh}$~\cite{Gieles:2008ew}. For isolated clusters $\xi_{\rm e}\simeq0.00739$, but $\xi_{\rm e}$ can be larger for systems experiencing the tidal field of the galaxy.
Along with the isolated cluster model, we also consider H\'enon's tidally limited model, for which $\xi_{\rm e}\simeq0.045$ and $\zeta\simeq0.0725$~\cite{Gieles:2011wh}: see Table~\ref{tab:cluster_evolutionary_scenarios}.

The tidally limited model can be applied to systems that experience the tidal field of the host galaxy within the inner kpc and inspiral to the center. On the other hand, the isolated model is more appropriate for the evolution of nuclear star clusters, since the tidal field cancels at the center of the galaxy. The isolated cluster evolution model is also adequate to describe clusters that reside several kpc away from the galaxy center; however, these isolated clusters in the halo of the galaxy are unlikely to merge with other clusters due to small coalescence cross sections.

\begin{table}
    \centering
    \begin{tabular}{ccc}
        \hline
         & $\zeta$ & $\xi_{\rm e}$ \\
        \hline
        isolated & 0.0926 & 0.00739 \\
        tidally limited & 0.0725 & 0.045 \\
        \hline
    \end{tabular}
    \caption{Parameters of the star cluster evolutionary scenarios we consider in this work, the isolated and tidally limited cases.} 
    \label{tab:cluster_evolutionary_scenarios}
\end{table}

\begin{figure*}
    \centering
    \includegraphics[width=\textwidth]{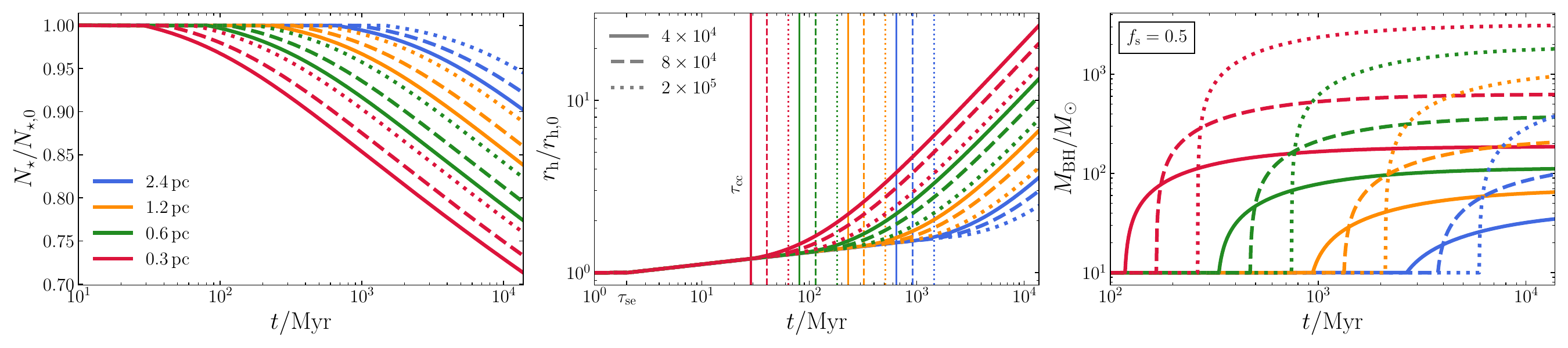}
    \includegraphics[width=\textwidth]{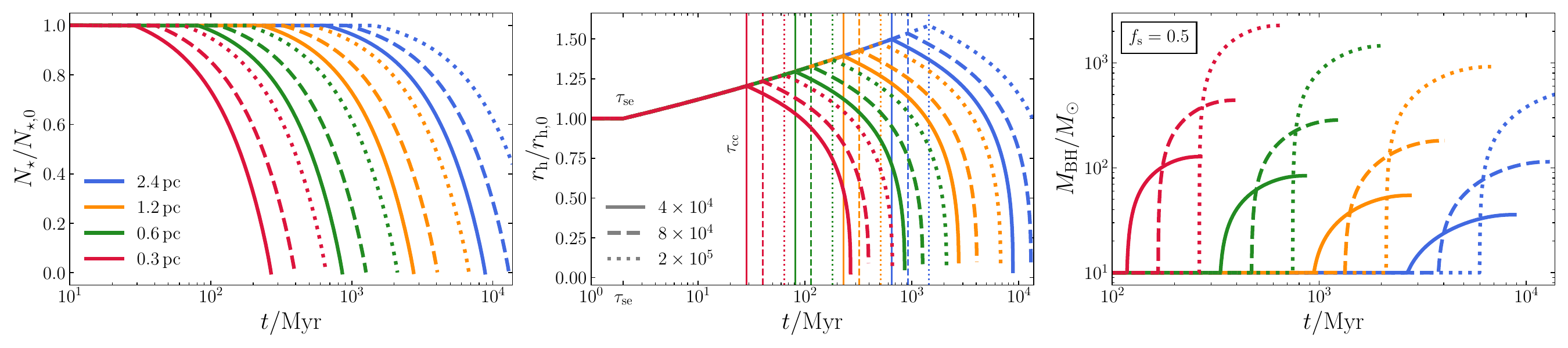}
    \caption{Time evolution of cluster mass (left panels), half-mass radius (middle panels), and black hole mass (right panels) for a grid of star cluster initial parameters $N_{\star,0}\in\{4\times10^4,8\times10^4,2\times10^5\}$ (corresponding to different line styles) and $r_{\rm h,0}\in\{0.3, 0.6, 1.2, 2.4\}\,\rm pc$ (corresponding to different colors). The top and bottom panels refer to isolated and tidally limited star clusters, respectively. We assume that a fraction $f_{\rm s}=0.5$ of each star's mass is consumed by the BH (see Sec.~\ref{Sec:Black_hole_mass_growth}). The evolution has been computed up to a maximum time $t$ corresponding to the Hubble time. The vertical lines in the middle panels mark the moment of core collapse and the start of balanced evolution  (i.e., $t=\tau_{\rm cc}$).}
    \label{Fig:Mcl_rh_MBH_t}
\end{figure*}

The system of equations for the global evolution is the following
\begin{subequations}
    \begin{align}
        {dN_{\star}\over dt}&=-\xi_{\rm e}{N_{\star}\over\tau_{\rm rh}}\Theta(t-\tau_{\rm cc})\,, \label{Eq:dMcldt} \\
        {dr_{\rm h}\over dt}&=(\zeta-2\xi_{\rm e}){r_{\rm h}\over\tau_{\rm rh}}\Theta(t-\tau_{\rm cc}) + {\nu r_{\rm h}\over t}\Theta(t - \tau_{\rm se})\,, \label{Eq:drhdt}
    \end{align}%
    \label{Eq:global_system}%
\end{subequations}
with initial conditions $N_\star(t=0)=N_{\star,0}$ and $r_{\rm h}(t=0)=r_{\rm h,0}$.
The mean stellar mass can easily be shown to evolve according to $\overline{m}_\star(t) = \overline{m}_{\star,0}(t/\tau_{\rm se})^{-\nu}$ for $t>\tau_{\rm se}$, with $\overline{m}_{\star,0}=0.6M_\odot$.
The mean mass quickly drops to below $0.4M_\odot$ within $1\,\rm Gyr$, after which it slowly continues to decrease.
Before the core-collapse time $t<\tau_{\rm cc}$, we assume that the global properties remain equal to the values set by the initial conditions, while the core collapses isothermally.
After $t=\tau_{\rm cc}$, gravothermal interactions produce heat in the center which is transferred throughout the cluster within $\sim\tau_{\rm rh}$.

Formally, the equation for $dN_\star/dt$ also contains the $-\Gamma_{\rm C}$ term, which is neglected here because its effect is subdominant since the flux of energy from tidal disruption of low angular momentum orbits is nearly zero~\cite{Merritt2013}; the cluster eventually expands in a self-regulatory way, such that the BH does not swallow the majority of the stars~\cite{1977ApJ...217..281S}.
For tidally limited systems, however, balanced evolution requires the cluster to contract because of the increased evaporation rate. 
In that case, the total stellar mass decreases to the point where the BH's contribution to the cluster's total mass becomes important and cannot be ignored.
Nevertheless, in Appendix~\ref{App:Cluster_dissolution_and_BH_feedback} we show that the omission of the $-\Gamma_{\rm C}$ term in the evolution of $N_\star$ and the approximation $M_{\rm cl}\simeq m_\star N_\star$ lead to a small relative error (of the order of $0.1\%$) on $M_{\rm BH}(t)$.
By neglecting this term and the contribution of $M_{\rm BH}$ to $M_{\rm cl}$, the equations for $N_{\star}$ and $r_{\rm h}$ separate from the evolution of $M_{\rm BH}$, and they can be solved independently.
Therefore, we first solve the two-by-two differential system for the global evolution of the cluster, then we plug in the solutions for $N_\star(t)$ and $r_{\rm h}(t)$ into the equation for $dM_{\rm BH}/dt$, and finally we integrate in time.

The system~\eqref{Eq:global_system} admits a simple analytic solution in closed form, which would not have been possible had we kept the $-\Gamma_{\rm C}$ term in the equation for $dN_\star/dt$.
To obtain it, we first divide Eq.~\eqref{Eq:dMcldt} by~\eqref{Eq:drhdt} to eliminate the relaxation time.
The resulting equation can be integrated to obtain a relation between $N_\star$ and $r_{\rm h}$,
\begin{align}
    {r_{\rm h}\over r_{\rm h,0}} = \left({\tau_{\rm cc}^2\over \tau_{\rm se}t}\right)^{\nu} \left({N_\star \over N_{\star,0}}\right)^{2 - \zeta/\xi_{\rm e}}.
    \label{Eq:Mcl_rh_relation}
\end{align}
Using this relation and the definition of $\tau_{\rm rh}$ to express~\eqref{Eq:drhdt} as an equation for $N_\star$, we find the relation
\begin{align}
    {dN_\star \over dt} = {-\xi_{\rm e}N_{\star,0}\over \tau_{\rm rh,0}} \left({t\over\tau_{\rm se}}\right)^{-{\nu/2}} \left({\tau_{\rm cc}^2\over \tau_{\rm se}t}\right)^{-3\nu/2}\left({N_\star \over N_{\star,0}}\right)^{3\zeta - 5\xi_{\rm e}\over2\xi_{\rm e}},
\end{align}
which can be integrated to get the time evolution of the stellar number
\begin{align}
    {N_\star(t)\over N_{\star,0}}=\begin{cases}
        1,\ t\le \tau_{\rm cc},\\
        \left[1 + {3\zeta - 7\xi_{\rm e}\over2\tau_{\rm rh,0}} {\tau_{\rm se}^{2\nu}\over \tau_{\rm cc}^{3\nu}} {(t^{\nu+1} - \tau_{\rm cc}^{\nu+1})\over\nu+1} \right]^{2\xi_{\rm e}\over7\xi_{\rm e}-3\zeta},\ t>\tau_{\rm cc}.
    \end{cases}
    \label{Eq:Nstart}
\end{align}
Substituting~\eqref{Eq:Nstart} into~\eqref{Eq:Mcl_rh_relation} then yields $r_{\rm h}(t)$.
When $\xi_{\rm e}=0$ and $\nu=0$ (i.e., in the case of no mass loss and no stellar evolution) this solution reduces to the late-time radius growth law $r_{\rm h}(t)\propto (t-\tau_{\rm cc})^{2/3}$: see, e.g., Eq.~(4) in~\cite{Antonini:2018auk}.

Above we have assumed that $\tau_{\rm se}<\tau_{\rm cc}$. The solutions should be modified when the core collapses on a timescale less than $\tau_{\rm se}=2\,\rm Myr$.
However, this requires extremely compact systems, which we do not consider here.
In those cases, stars would rapidly condense in the system's center and stellar collisions would efficiently produce a massive stellar runaway~\cite{PortegiesZwart:2002iks}.

In the tidally limited case, a massive cluster may dissolve in a finite time. The evaporation condition is
\begin{equation}
N_\star(t=\tau_{\rm ev})=0,
\label{eq:tau_evap}
\end{equation}
which can be solved numerically to obtain the evaporation time $\tau_{\rm ev}$. In particular, the equation has a real positive root if and only if $7\xi_{\rm e} > 3\zeta$.

In Fig.~\ref{Fig:Mcl_rh_MBH_t} we show the time evolution of the stellar number (left panels) and half-mass radius (middle panels) for a set of initial conditions.
The top panels show the evolution of isolated clusters, while the bottom panels show the evolution of tidally limited clusters.
The evaporation rate (and hence the expansion/contraction rate) is more sensitive to the compactness of the cluster than to its mass, because the half-mass relaxation time has a stronger dependence on the radius than on the mass: $\tau_{\rm rh}\propto \sqrt{N_\star r_{\rm h}^{3}}$.
The slow growth of $r_{\rm h}$ visible at $2\,\rm Myr$ in the central panels, before balanced evolution begins (at $\tau_{\rm cc}$) for all models,
corresponds to the adiabatic expansion due to stellar wind mass loss, and it can be attributed to the last term of Eq.~\eqref{Eq:radius_evolution_equation}.

\subsection{Black hole mass growth}
\label{Sec:Black_hole_mass_growth}

Having computed the global evolution of the cluster, we can now find the growth of the BH mass as a function of time by integration.

While the BH subsystem remains Spitzer unstable, it decouples from stars and evolves independently.
The formation of hard binary BHs happens in the central regions, and binary-single BH interactions often lead to ejections.
During this balanced evolutionary phase, the hard binary BHs generate the energy required for the cluster to expand.
After most BHs have evaporated from the system and the stellar population starts to couple with the few remaining BHs dynamically, the heat production in the system becomes dominated by the bounded stars within the influence radius of the BH~\cite{Rizzuto:2022fdp}.
Then the runaway tidal interactions begin, and a single growing BH stands out and settles in the center.
According to Ref.~\cite{Breen:2013vla} it takes about 10 initial relaxation times after core collapse for 90\% of the BHs to be ejected from the cluster; that is, the lifetime of the BH subsystem is $\tau_{\rm BH}\sim\tau_{\rm cc}+10\tau_{\rm rh,0}\simeq13.21\tau_{\rm rh,0}$.
We define this timescale to correspond to the moment when the BH seed begins to grow via runaway tidal encounters with stars.

We begin with $M_{\rm BH,0}=10M_\odot$ stellar-mass BH seeds in star clusters and follow their growth through the repeated consumption of stars using the rate formula for $\Gamma_{\rm C}$ in Eq.~\eqref{Eq:consumption_rate}. 
We have checked that taking $M_{\rm BH,0}=50M_\odot$ has only a mild effect on the growth history of the BH, leading to mass variations with a relative error of order less than 5\% (see also Ref.~\cite{Antonini:2018auk}). This implies a weak dependence on metallicity. Thus, our results are rather insensitive to the seed BH mass spectrum, as long as the seeds are of stellar mass. Assuming that a fraction $f_{\rm s}$ of each star's mass is consumed by the BH, the mass accretion rate can be expressed as
\begin{equation}
  {d{M}_{\rm BH}\over dt}=f_{\rm s}\overline{m}_\star \Gamma_{\rm C}
  \label{eq:dMBHdt}
\end{equation}
for $t>\tau_{\rm BH}$.
Using the definition of $\Gamma_{\rm C}$ from Eq.~\eqref{Eq:consumption_rate}, we find
\begin{align}
    {dM_{\rm BH}\over dt}\simeq3.4f_{\rm s}R_\star\sqrt{G}\overline{m}_\star^{13/6}\left({N_\star\over r_{\rm h}}\right)^{5/2}{1\over M_{\rm BH}^{2/3}}\,,
    \label{Eq:dMBHdt}
\end{align}
where $R_\star$ is the radius of the star (for which we use the solar value).
Eq.~\eqref{Eq:dMBHdt} can be integrated by separation of variables, and the BH mass evolution is found to be
\begin{widetext}
    \begin{align}
    M_{\rm BH}(t) = \begin{cases}
        M_{\rm BH,0},\ t\le\tau_{\rm BH},\\
        \left( M_{\rm BH,0}^{5/3} + c_1\left[ x^{{\nu\over3}+1} \prescript{}{2}F_{1}\left( {\nu+3\over3(\nu+1)}, {5(\xi_{\rm e}-\zeta)\over7\xi_{\rm e} - 3\zeta}; {4\nu + 6\over3(\nu+1)};  \left({x\over \tau_{\rm cc}}\right)^{\nu+1} {1\over 1 - {2(\nu+1)\over3\zeta - 7\xi_{\rm e}} {\tau_{\rm rh,0}\over\tau_{\rm cc}} \left({\tau_{\rm cc}\over \tau_{\rm se}}\right)^{2\nu}}  \right) \right]\Bigg|_{\tau_{\rm BH}}^t  \right)^{3/5},\ t>\tau_{\rm BH}
    \end{cases}
    \label{Eq:BH_mass_growth_equation}
    \end{align}
\end{widetext}
where $x$ is the integration variable, $\prescript{}{2}F_{1}$ is the hypergeometric function, and we have defined the constant
    \begin{align}
      c_1 &= \sqrt{50G\kappa^{5}\over2187\pi}{3-\gamma\over2-\gamma}{\Gamma(\gamma+1)\over\Gamma(\gamma-1/2)}\nonumber\\
      &\times f_{\rm s}\beta \eta^{2/3}R_\star \overline{m}_{\star,0}\left({N_{\star,0}\over r_{\rm h,0}}\right)^{5/2} {\tau_{\rm se}^{14\nu/3}\over \tau_{\rm cc}^{5\nu}}\,,
    \end{align}
where $\Gamma$ denotes the gamma function.
We can invert Eq.~\eqref{Eq:BH_mass_growth_equation} numerically (in practice, we use a Newton-Raphson scheme) to obtain the time $t$ at which the BH mass is equal to $M_{\rm BH}$ with a tolerance of at most $1\,\rm Myr$.

At late times, the BH mass asymptotes to a constant value that depends on the initial conditions of the cluster, and correlates more strongly with the initial cluster mass than with the initial half-mass radius.
In general, heavier star clusters produce heavier IMBHs. By contrast, the half-mass radius mostly affects the relaxation time, and hence the time at which the BH starts growing.
The evolution of the BH mass as a function of time is shown in the rightmost panels of Fig.~\ref{Fig:Mcl_rh_MBH_t}.
In the tidally limited case (lower right panel) the BH mass curve terminates at the moment the cluster evaporates, defined in Eq.~\eqref{eq:tau_evap}.

\subsection{Dependence on the cluster's initial conditions}
\label{sec:Dependence_on_cluster's_initial_condition}

\begin{figure*}
    \centering
    \includegraphics[width=0.49\textwidth]{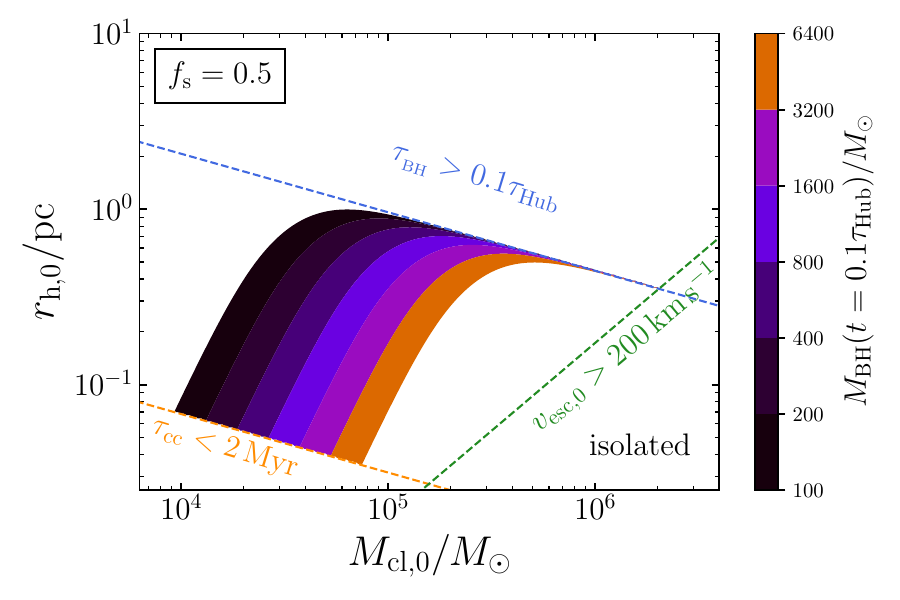}
    \includegraphics[width=0.49\textwidth]{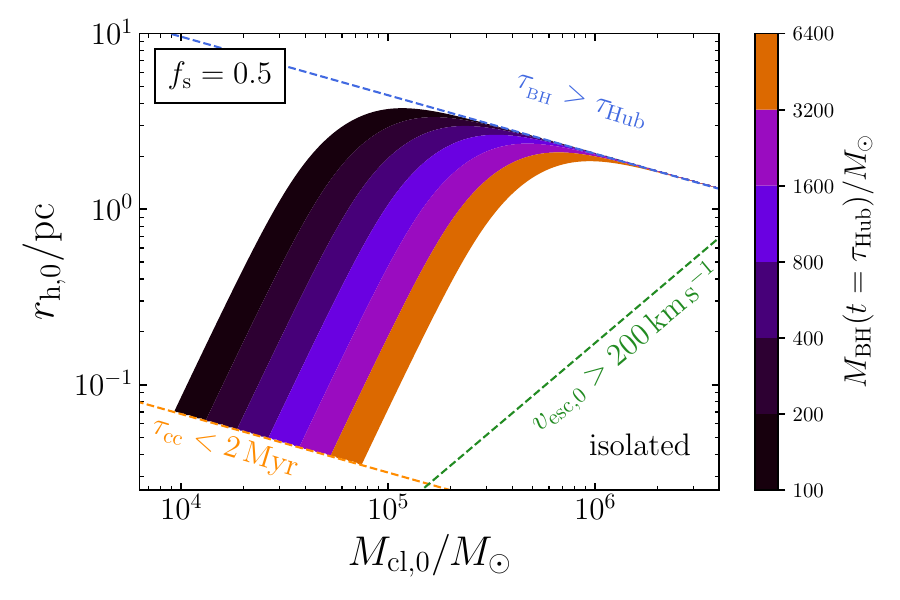}
    \includegraphics[width=0.49\textwidth]{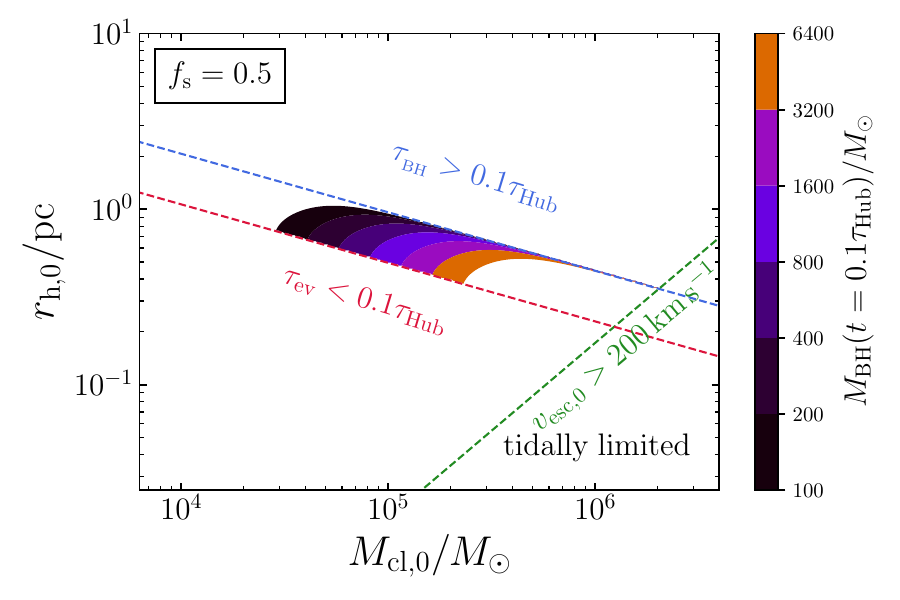}
    \includegraphics[width=0.49\textwidth]{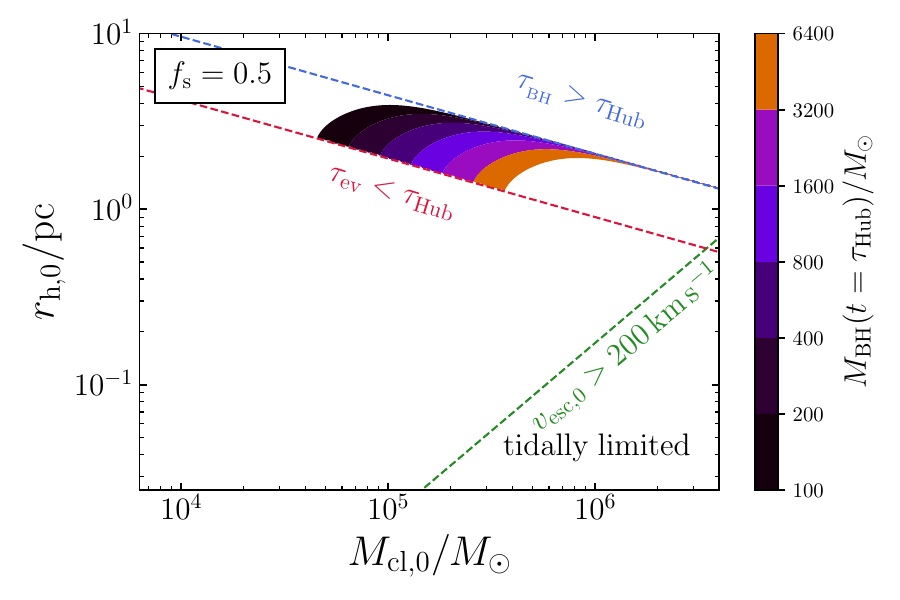}
    \caption{Contour plots of the BH mass assembled via runaway tidal encounters in the $(M_{\rm cl,0},r_{\rm h,0})$ plane at $t=0.1\tau_{\rm Hub}$ (left panels) and $t=\tau_{\rm Hub}$ (right panels). The top and bottom panels refer to isolated and tidally limited star clusters, respectively. The $\tau_{\rm BH}\ge\tau_{\rm Hub}$ ($\tau_{\rm BH}\ge0.1\,\tau_{\rm Hub}$) region above the blue dashed line corresponds to clusters for which the time required for the BH to begin growing is larger than a Hubble time (one-tenth of a Hubble time) due to nonevaporation of the BH subsystem. In the top panels, the $\tau_{\rm cc}\le 2\,\rm Myr$ region below the dashed orange line corresponds to clusters for which core collapse occurs before massive stars become BHs, and an IMBH might form through runaway stellar mergers instead.
      The $v_{\rm esc,0}>200\,\rm km\, s^{-1}$ region (dashed green line) corresponds to clusters for which the initial escape velocity is large enough for the repeated BH merger channel to dominate. In the bottom panels, the $\tau_{\rm ev}\le\tau_{\rm Hub}$ ($\tau_{\rm ev}\le0.1\tau_{\rm Hub}$) region corresponds to clusters that evaporate on a timescale less than $\tau_{\rm Hub}$ ($0.1\tau_{\rm Hub}$): see Eq.~\eqref{eq:tau_evap}. The parameter $f_{\rm s}$ is set to 0.5.}
    \label{Fig:Mcl0_rh0_MBH_}
\end{figure*}

\begin{figure}
    \centering
    \includegraphics[width=0.49\textwidth]{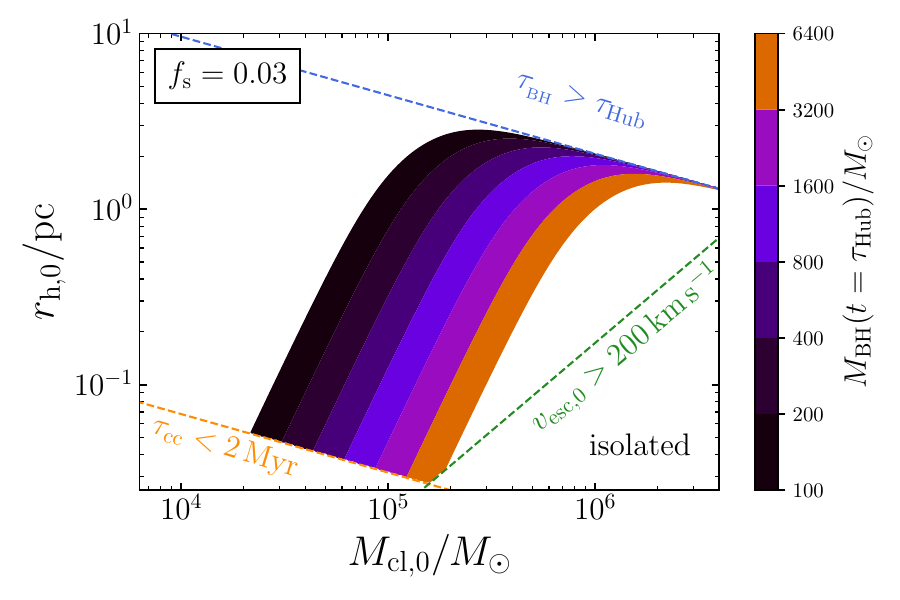}
    \caption{Same the as the top right panel of Fig.~\ref{Fig:Mcl0_rh0_MBH_}, but with $f_{\rm s}=0.03$.}
    \label{Fig:Mcl0_rh0_MBH_fs003}
\end{figure}

\begin{figure*}
    \centering
    \includegraphics[width=0.49\textwidth]{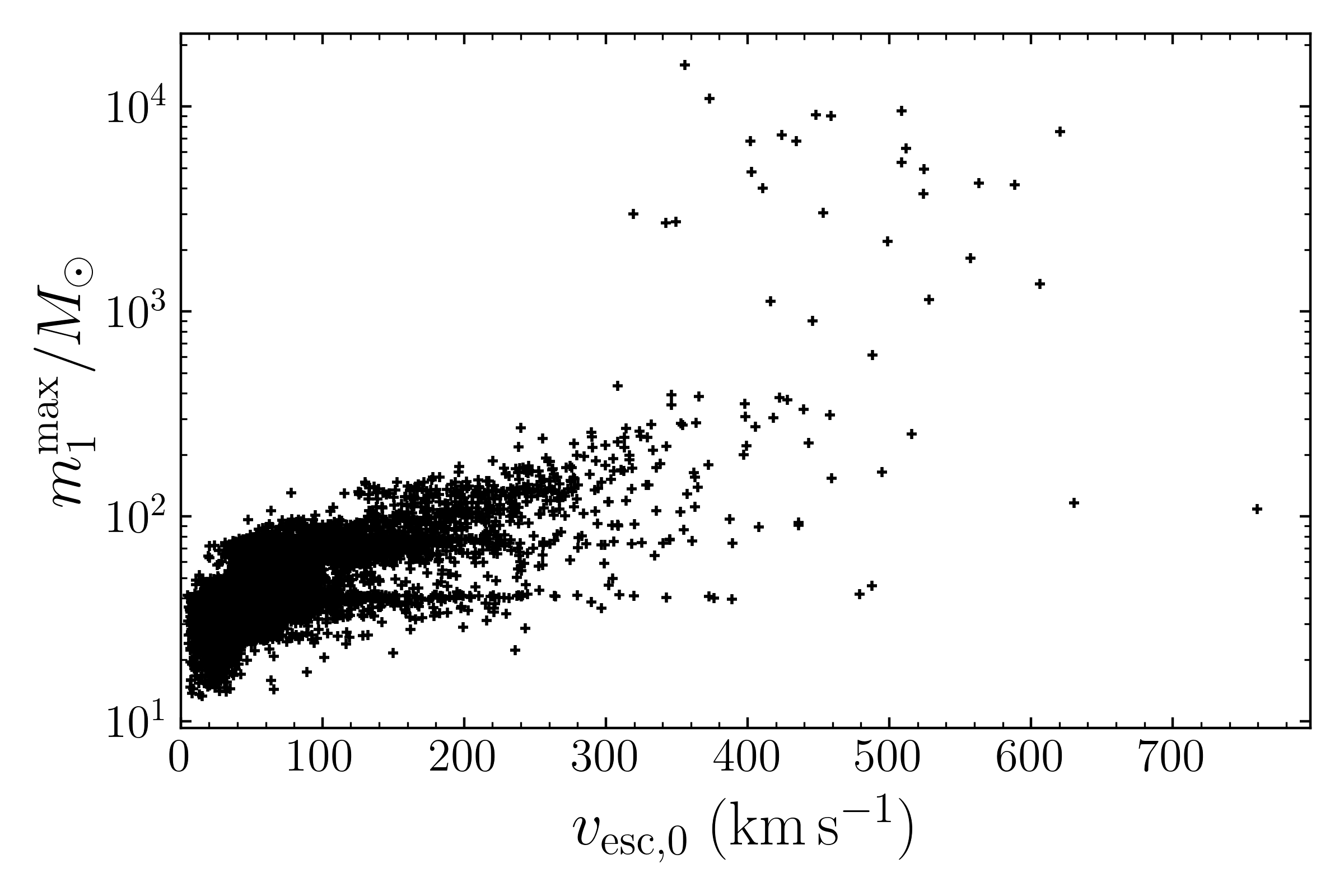}
    \includegraphics[width=0.49\textwidth]{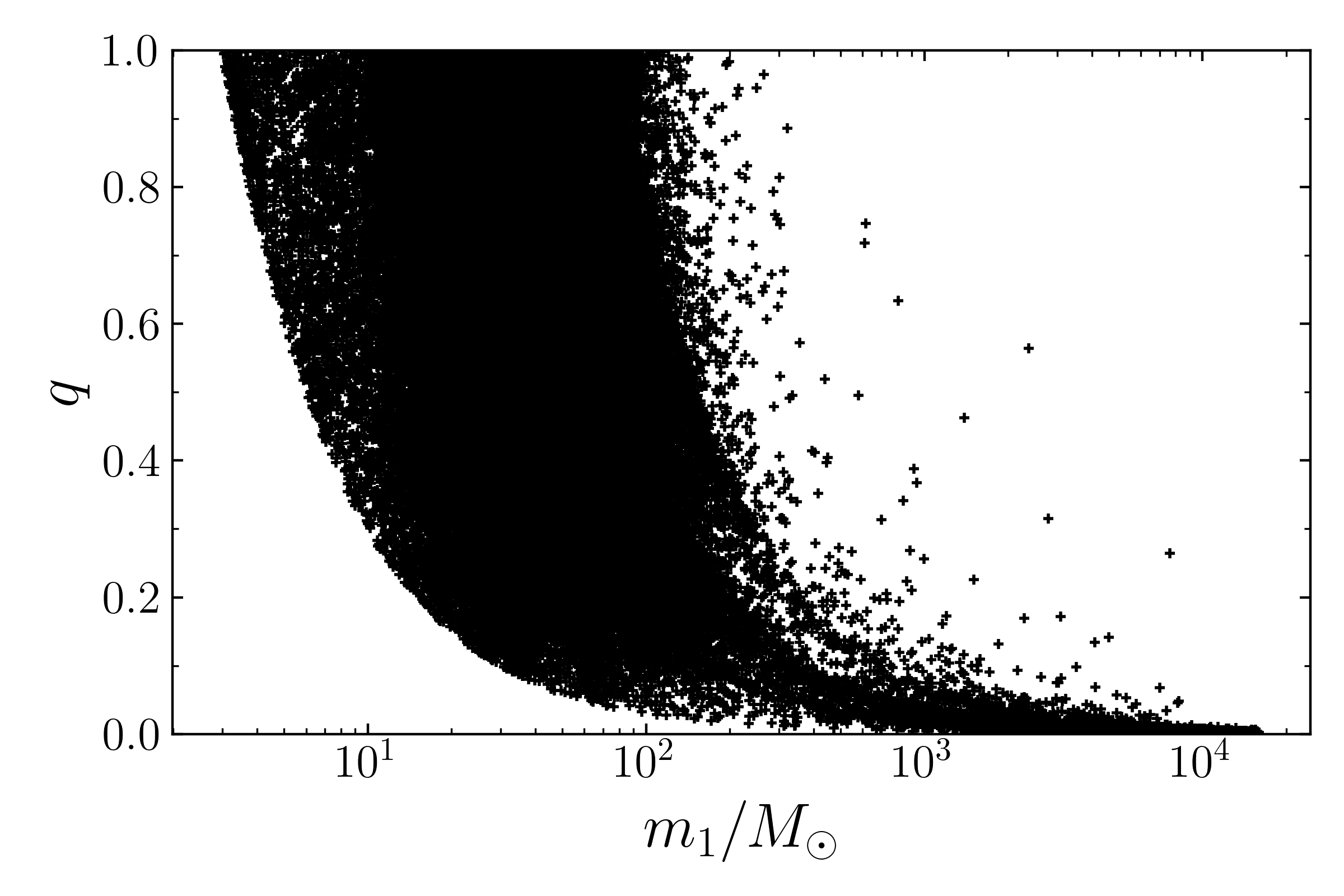}
    \caption{Left panel: maximum BH mass assembled through repeated BH mergers in a large set of {\sc Rapster} simulations as a function of the initial cluster escape velocity. Each point corresponds to a different star cluster simulation. Right panel: mass ratio versus the primary mass of each merger in the {\sc Rapster} catalog.}
    \label{Fig:vesc_mbhmax__m1_q}
\end{figure*}

The fraction of stellar mass accreted, $f_{\rm s}$, can be as high as $\sim 0.5$~\cite{1988Natur.333..523R} and as low as $\sim 0.01$~\cite{Metzger:2015sea}.
In this work, we consider two representative values of $f_{\rm s}=0.5$ and $f_{\rm s}=0.03$.
In Fig.~\ref{Fig:Mcl0_rh0_MBH_} we set $f_{\rm s}=0.5$, and we show the BH mass assembled at $t=0.1\,\tau_{\rm Hub}$ (left panel) and $t=\tau_{\rm Hub}$ (right panel), where $\tau_{\rm Hub}$ is the Hubble time, in the $(M_{\rm cl,0},\, r_{\rm h,0})$ plane, where $M_{\rm cl,0}$ is the initial cluster mass and $r_{\rm h,0}$ is the initial half-mass radius. The top panels correspond to the isolated evolutionary scenario, while the bottom panels correspond to the tidally limited case. In Fig.~\ref{Fig:Mcl0_rh0_MBH_fs003} we illustrate the effect of a lower value of $f_{\rm s}$ (here, $f_{\rm s}=0.03$) by plotting the BH mass assembled at $t=\tau_{\rm Hub}$ in the isolated case (the other panels have been omitted for brevity, as they follow similar trends). By comparing with the top right panel of Fig.~\ref{Fig:Mcl0_rh0_MBH_}, we see that a lower value of $f_{\rm s}$ results in a shift of the constant BH mass contours: for a fixed $M_{\rm cl,0}$ the BH mass is smaller, and a heavier cluster is required to compensate for the loss in the stellar mass accreted at each TDE.

In the top panels, the region below the orange dashed lines corresponds to systems whose core collapses on a timescale less than $2\,\rm Myr$ and runaway stellar collisions would form a supermassive star that collapses into an IMBH~\cite{2018MNRAS.478.2461G}. We thus exclude this region ($\tau_{\rm cc}<2\,\rm Myr$) of the parameter space since it produces massive BHs from another channel.
In the bottom panels, the $\tau_{\rm ev}\le\tau_{\rm Hub}$ ($\tau_{\rm ev}\le0.1\tau_{\rm Hub}$) region corresponds to clusters that evaporate on a timescale less than $\tau_{\rm Hub}$ ($0.1\tau_{\rm Hub}$), where we recall that the criterion for cluster evaporation was given in Eq.~\eqref{eq:tau_evap}.

The green dashed line corresponds to a constant initial escape velocity of $200\,\rm km\, s^{-1}$~\cite{Antonini:2018auk}. 
We exclude the region to the right of this line from this study, because when the escape velocity is large enough, it is possible to have a significant number of repeated BH mergers and to form an IMBH in the core of the cluster while the BH subsystem has not evaporated yet. In fact, the main limiting factor in growing BHs through repeated mergers comes from the GW recoil received by the remnant of each merger event, which can often of order hundreds of $\rm km\, s^{-1}$. For ths reason, BH retention requires environments that are sufficiently heavy and compact~\cite{Gerosa:2019zmo}. 
A BH growing through repeated mergers can be efficiently retained in clusters as long as the escape velocities are larger than $\approx300\, \rm km\, s^{-1}$, leading to a runaway growth~\cite{Zevin:2022bfa}. 
Reference~\cite{Antonini:2018auk} also finds that the formation of IMBHs above $100M_\odot$ is efficient if the escape velocity exceeds a critical value which is in the range $\simeq200$--$300\,\rm km\, s^{-1}$. 

To demonstrate the significance of repeated BH mergers, we have performed a set of star cluster simulations with the {\sc Rapster} code~\cite{Kritos:2022ggc} (for the details of our initial conditions on the simulated star cluster population, see Sec.~III.A of Ref.~\cite{Yi:2024elj}).
In the left panel of Fig.~\ref{Fig:vesc_mbhmax__m1_q} we show the maximum primary BH mass assembled in each simulated cluster, while in the right panel we show the mass ratio of every dynamical merger from all cluster simulations.
The maximum BH mass is strongly correlated with the initial escape velocity: heavier BHs form in systems with higher escape velocity, because the GW kick received by the remnant limits the retention of the runaway BH.
The simulations indicate that hierarchical BH mergers contaminate the BH mass spectrum above $200M_\odot$ if the cluster's initial escape velocity is in the hundreds of $\rm km\, s^{-1}$.
Moreover, as can be seen from the right panel of Fig.~\ref{Fig:vesc_mbhmax__m1_q}, the growth of BHs through repeated mergers is predominantly oligarchic, i.e., the probability of forming two massive BHs in the same system is very low.
If two moderately heavy BHs form in the same cluster, then the two heaviest BHs will preferentially pair up and merge before capturing other smaller BHs.
In particular, in the primary mass range above $200M_\odot$ the mass ratio typically remains below $\sim 0.3$.
In this way it is possible to form binaries involving two massive BHs from the merger of two different stellar systems, each of which formed an IMBH.

We emphasize that the exact location of the decision boundaries marked by the orange, red, and blue dashed lines in our parameter space are qualitative features, motivated by the discussion above. 

\subsection{Delay time}
\label{Sec:Delay_time}

We consider the formation of IMBH-IMBH pairs through the coalescence of two star clusters, each cluster hosting one IMBH (see Fig.~\ref{Fig:parameter_description}). 
This can be accomplished either through the merger of globular clusters that inspiral into the central regions of their galaxy through dynamical friction~\cite{Ebisuzaki:2001qm} or through the coalescence of nuclear star clusters in the postmerger phase of two dwarf galaxies~\cite{Palmese:2020xmk}. 
Alternatively, IMBHs may form in star clusters that fragment in the turbulent central regions of a galaxy and are then brought together to form a bound pair~\cite{Shi:2024wgh}. 

To compute the new properties of the merged star cluster, we assume mass and energy conservation (see also Ref.~\cite{Lupi:2014vza}), so that if $M_{\rm cl,1}$ and $M_{\rm cl,2}$ are the masses and $r_{\rm h,1}$ and $r_{\rm h,2}$ the half-mass radii of the two clusters just before their coalescence, then the mass and radius of the new cluster are given by
\begin{subequations}
\begin{align}
    M_{\rm cl}' &= M_{\rm cl,1} + M_{\rm cl,2}\,,\\
    r_{\rm h}'&= {M_{\rm cl}'^2\over {M_{\rm cl,1}^2\over r_{\rm h,1}} + {M_{\rm cl,2}^2\over r_{\rm h,2}}}\,.
\end{align}
\end{subequations}

The merger of the two IMBHs brought into the same environment is not instantaneous. Rather, there is a time delay $\tau_{\rm delay}$ following the merger of the two clusters.
We thus identify three contributions to the time delay:
\begin{itemize}
    \item[(1)] The dynamical friction timescale over which the least massive IMBH with mass $m_2$ reaches the center from the initial half-mass radius of the remnant cluster $r_{\rm h}'$. This is the time for the formation of the IMBH-IMBH binary since the merger of their host clusters. It is given by (see Ref.~\cite{1988gady.book.....B}, pg.~648)
    \begin{align}
        \tau_{\rm df}\simeq3.8\,{\rm Myr}\, \left({r_{\rm h}'\over1\,\rm pc}\right)^2 {\sigma'\over10{\,\rm km\, s}^{-1}}{100M_\odot\over m_2},
    \end{align}
    where we have set the Coulomb logarithm to $\ln\Lambda=10$ and $\sigma'=(0.4GM_{\rm cl}'/r_{\rm h}')^{1/2}$. Since $\tau_{\rm df}$ scales inversely with the IMBH mass, the heaviest IMBH reaches the center first, thus the formation time of the IMBH-IMBH binary is dominated by the dynamical friction of the secondary IMBH.
    
    \item[(2)] The hardening timescale of the IMBH-IMBH binary, $\tau_{\rm har}$. 
    We compute the hardening timescale by assuming that the pair hardens through binary-single interactions at a rate provided by full-loss cone theory. Stars that interact closely with the hard IMBH-IMBH binary are ejected and are unable to efficiently heat up the cluster. 
    
    If $E_{\rm b}=Gm_1m_2/(2a)$ is the binding energy of the IMBH-IMBH binary, then the rate of change of binding energy is given by $dE_{\rm b}/dt=\langle\Delta E_{\rm b}\rangle (dN/dt)$. Here the mean energy change per interaction is given by $\langle\Delta E_{\rm b}\rangle=(H/(4\pi))(m_\star/m_{12})E_{\rm b}$, where $H\simeq 15$ and $m_{12}=m_1+m_2$~\cite{Quinlan:1996vp}.
    Moreover, $dN/dt$ is the binary-single interaction rate given by Eq.~(6.15) of Ref.~\cite{Merritt:2003pf}, computed by setting the loss-cone radius equal to the semimajor axis of the binary, $a_{\rm B}$, and the central mass equal to the binary's mass, $m_{12}$. Combining these equations leads to a differential equation for the evolution of $a_{\rm B}$
    \begin{align}
        {da_{\rm B}\over dt}=-{H\over2\pi}\sqrt{2\over\pi}{3-\gamma\over2-\gamma}{\Gamma(\gamma+1)\over\Gamma(\gamma-1/2)} {\sigma'^5\over G^2m_{12}^2}a_{\rm B}^2.
        \label{Eq:daBdt}
    \end{align}
    In addition, we assume a Bahcall-Wolf profile, i.e., $\gamma=7/4$. For simplicity, we further assume that the cluster properties do not vary significantly over the hardening evolution of the IMBH-IMBH binary. 
    Therefore we approximate $\sigma'$ as constant during the evolution of the IMBH-IMBH pair until its merger. 
    
    Based on a cross-section argument (for the derivation see, e.g., Sec.~4.3 from Ref.~\cite{Celoria:2018mzr}), the rate of change of the binary's semimajor axis can be written as $da_{\rm B}/dt = -(GH\rho/\sigma')a_{\rm B}^2$. Notice that Eq.~\eqref{Eq:daBdt} can be brought into this form if we define an effective density as
    \begin{align}
        \rho_{\rm eff}\equiv{1\over2\pi}\sqrt{2\over\pi}{3-\gamma\over2-\gamma}{\Gamma(\gamma+1)\over\Gamma(\gamma-1/2)} {\sigma'^6\over G^3m_{12}^2}.
    \end{align}
    The semimajor axis shrinks until the emission of GWs is efficient enough to drive the evolution of the binary. At this point, the binary decouples from the dynamics, and we denote the corresponding semimajor axis by $a_{\rm gw}$.
    The eccentricity at decoupling, $e_{\rm gw}$, is sampled from a thermal distribution, and the hardening timescale $\tau_{\rm har}$ is computed as in Eq.~(9) from Ref.~\cite{Sesana:2015haa}, where the density is set equal to $\rho_{\rm eff}$.
    \item[(3)] The gravitational radiation coalescence timescale, $\tau_{\rm gw}$. We compute this timescale by substituting the semimajor axis and the eccentricity of the binary at dynamical decoupling, $a_{\rm gw}$ and $e_{\rm gw}$, respectively, as computed in the previous step, in the semianalytic approximation formula from Ref.~\cite{Mandel:2021fra}.
\end{itemize}
To summarize, the total time delay is computed as $\tau_{\rm delay}=\tau_{\rm df} + \tau_{\rm har} + \tau_{\rm gw}$.

\section{Single-event cluster posteriors}
\label{sec:Single-event_cluster_posteriors}

We perform full Bayesian PE on six IMBH-IMBH merging systems, denoted in alphabetical order from A to F. Their parameters are listed in Table~\ref{tab:binary_parameters}. We assume a network of three XG GW detectors, including two CEs and one ET. In Appendix~\ref{App:Parameter_estimation_of_IMBH-IMBH_binaries} we give more details on the PE runs, the choice of network, and the chosen waveform model. 
We set the spin magnitude of all IMBHs to a small but nonzero
value of $0.1$.
This choice is motivated by the expected evolution of the IMBH spin as the mass grows through repeated minor TDEs: as discussed in Appendix~\ref{App:mass_spin_evolution}, the spin rapidly asymptotes to zero as the BH mass increases, with a scatter of order $\sim0.1$. 

For each IMBH-IMBH binary, we exploit the PE results and the forward model discussed in Sec.~\ref{sec:Bayesian_analysis} below to hierarchically infer the initial properties of the cluster environment in which the IMBHs formed. The Bayesian framework to compute the cluster posterior distributions and our main results are presented in Sec.~\ref{sec:Bayesian_analysis} and Sec.~\ref{sec:Results}, respectively. 

\begin{table}
    \centering
    \begin{tabular}{ccccc}
        \hline
          & $m_1/M_\odot$ & $m_2/M_\odot$ & $z_{\rm m}$ & SNR \\
        \hline
         A & 300 & 300 & 1.0 & 515 \\
         B & 300 & 300 & 2.0 & 211 \\
         C & 300 & 300 & 4.0 & 64 \\
         D & 600 & 200 & 2.0 & 225 \\
         E & 1000 & 400 & 0.5 & 1548 \\
         F & 200 & 200 & 8.0 & 17 \\
         \hline
    \end{tabular}
    \caption{Source parameters of the six IMBH-IMBH binaries on which we perform PE. The columns are from left to right: primary (source-frame) IMBH mass, secondary (source-frame) IMBH mass, merger redshift, and network SNR.}
    \label{tab:binary_parameters}
\end{table}

\subsection{Bayesian analysis}
\label{sec:Bayesian_analysis}

In this subsection we present the Bayesian framework used to infer the cluster properties. We start with a description of the cluster parameters (Sec.~\ref{sec:Parameter_description}), followed by the method we use to reweigh the prior (Sec.~\ref{sec:reweighing}) and a discussion of our chosen astrophysical prior (Sec.~\ref{sec:Astrophysical_prior}).

\subsubsection{Cluster parameters}
\label{sec:Parameter_description}

Our goal is to obtain the posterior distribution of the cluster parameters $\boldsymbol{\lambda}$ given the GW data $d$
corresponding to a single IMBH-IMBH merger event, which we denote by
$p(\boldsymbol{\lambda}|d)$.  
We interpret $\boldsymbol{\lambda}$ as the collection of initial conditions of the parent clusters associated with the two IMBHs in the merger event, i.e.,
\begin{align}
    \boldsymbol{\lambda} = \{ & M_{\rm cl,01}, r_{\rm h,01}, z_{\rm cl,01}, \Delta t_{\rm g,1}, \nonumber\\ & M_{\rm cl,02}, r_{\rm h,02}, z_{\rm cl,02}, \Delta t_{\rm g,2} \}.
    \label{Eq:hyper-parameters}
\end{align}
Recall that $\Delta t_{\rm g,i}$ ($i=1,2$) denotes the growth time of each cluster, i.e., the time elapsed between the formation of each cluster and their merger. 
We want to infer the cluster parameters from information about the IMBH component masses and their merger redshift $\boldsymbol{\theta}=\{m_1, m_2, z_{\rm m}\}$. 
The analytical model of Sec.~\ref{sec:Black_hole_growth_model} relates $\boldsymbol{\theta}$ and $\boldsymbol{\lambda}$ through a map of the form ${\cal F}:\boldsymbol{\lambda}\to\boldsymbol{\theta}$. 
More formally, we have
\begin{subequations}
    \begin{align}
    m_1 &= M_{\rm BH}(\Delta t_{\rm g,1}; M_{\rm cl,01}, r_{\rm h,01})\,,\label{Eq:m1=MBH1} \\
    m_2 &= M_{\rm BH}(\Delta t_{\rm g,2}; M_{\rm cl,02}, r_{\rm h,02})\,,\label{Eq:m2=MBH2} \\
    z_{\rm m} &= z(t_{\rm cl,01} + \Delta t_{\rm g,1} + \tau_{\rm delay}) \nonumber\\&= z(t_{\rm cl,02} + \Delta t_{\rm g,2} + \tau_{\rm delay})\,. \label{Eq:zm}
\end{align}%
\label{Eq:F_map}%
\end{subequations}%
Here $z=z(t)$ is the redshift-cosmic time relation computed assuming the Planck 2018 cosmology~\cite{Planck:2018vyg}, and $M_{\rm BH}$ is the BH mass growth function of Eq.~\eqref{Eq:BH_mass_growth_equation}. We have also made explicit the dependence on the initial mass and half-mass radius of each cluster. 
We impose the time constraint
\begin{equation}
t_{\rm cl,01} + \Delta t_{\rm g,1} = t_{\rm cl,02} + \Delta t_{\rm g,2}\,,
\label{eq:time_constraint}
\end{equation}
where $t_{\rm cl,0i}$ are the cosmic times of cluster formation corresponding to redshifts $z_{\rm cl,0i}$ ($i=1,2$), because the two clusters must merge at the same cosmic time. 
In Eq.~\eqref{Eq:zm}, the delay time $\tau_{\rm delay}$ (that depends on $\boldsymbol{\lambda}$) is computed as the sum of three time delays as described in Sec.~\ref{Sec:Delay_time}. 
We do not include $\tau_{\rm delay}$ in the list of Eq.~\eqref{Eq:hyper-parameters} because it is not independent from the other parameters: $\tau_{\rm delay}$ must respect the condition $t_{\rm cl,01} + \Delta t_{\rm g,1} + \tau_{\rm delay} = t_{\rm cl,02} + \Delta t_{\rm g,2} + \tau_{\rm delay} = t_{\rm m}$, which is equivalent to Eq.~\eqref{Eq:zm}, because the IMBHs must merge at the same cosmic time $t_{\rm m}$ (or at the same redshift $z_{\rm m}$). 

Conservation of probability implies that
$p(\boldsymbol{\lambda}|d)d\boldsymbol{\lambda} = p(\boldsymbol{\theta}|d)d\boldsymbol{\theta}$. 
Therefore, given posterior samples $\{\boldsymbol{\theta}_i\}$ of the source parameters of a GW event, we may compute posterior samples of the cluster parameters $\{\boldsymbol{\lambda}_i\}$ by using the inverse map $\boldsymbol{\lambda}_i={\cal F}^{-1}(\boldsymbol{\theta}_i)$. 
Notice that the $\cal F$-map is many-to-one as it is a projection, and thus the value of its inverse is not unique. 
Nevertheless, we still consider an inverse scheme in which we randomly sample points from the five-dimensional submanifold ${\cal F}^{-1}(\boldsymbol{\theta})$.
In practice, we first sample $\log_{10}(M_{\rm cl,0})$ and $\log_{10}(r_{\rm h,0})$ for both clusters uniformly and compute $\Delta t_{\rm g,1}$ and $\Delta t_{\rm g,2}$ from Eq.~\eqref{Eq:m1=MBH1} and~\eqref{Eq:m2=MBH2}, respectively, subject to the constraints that (i) the inferred growth times are positive definite; (ii) the lifetime of each BH subsystem is $\tau_{\rm BH}<\tau_{\rm Hub}$; (iii) each cluster needs to collapse after stars become BHs, i.e., $\tau_{\rm cc}>\tau_{\rm se}$; (iv) each escape velocity needs to be less than the critical escape velocity, i.e., $v_{\rm esc,0}<200\,\rm km\, s^{-1}$; and finally (v) the age of each cluster formation $t_{\rm cl,0}$, calculated from Eqs.~\eqref{Eq:zm} and~\eqref{eq:time_constraint}, must be larger than a minimum value $\tau_{\rm PopII}$, here taken to correspond to a redshift $z_{\rm PopII}=15$.

Our sampling algorithm of the submanifold ${\cal F}^{-1}(\boldsymbol{\theta})$ does not return a uniform set of draws. Therefore, we reweigh our samples with the inverse Jacobian factor $|d\boldsymbol{\theta}/d\boldsymbol{\lambda}|^{-1}$, which we derive to be the product of the TDE rates [see Eq.~\eqref{Eq:dMBHdt}] at the moment of cluster merger. To derive this weight factor, we define a new map $\boldsymbol{\theta}'\to\lambda$ with $\boldsymbol{\theta}'=\{ m_1, m_2, z_{\rm m}, z_{\rm m}', M_{\rm cl,01}', M_{\rm cl,02}', r_{\rm h,01}' r_{\rm h,02}' \}$, where $m_1$, $m_2$, and $z_{\rm m}$ are defined as in Eq.~\eqref{Eq:F_map}, $z_{\rm m}'=z_{\rm m}$, $M_{\rm cl,0i}'=M_{\rm cl,0i}$, and $r_{\rm h,0i}'=r_{\rm h,0i}$ ($i=1,2$). We then compute the determinant of the Jacobian matrix for this map and obtain the product of the inverse of the TDE rates evaluated at the moment of cluster merger. 

\subsubsection{Source-parameter prior reweighing}
\label{sec:reweighing}

When performing PE on our chosen events, we assume some prior on the source parameters, denoted by $p_{\rm PE}(\boldsymbol{\theta})$. 
Since the likelihood function remains invariant under the $\cal F$-map, so that
$p(d|\boldsymbol{\theta})=p(d|\boldsymbol{\lambda})$,
it follows that the priors on $\boldsymbol{\theta}$ and $\boldsymbol{\lambda}$ are related by a Jacobian as $p_{\rm PE}(\boldsymbol{\theta})d\boldsymbol{\theta} = p_{\rm PE}(\boldsymbol{\lambda})d\boldsymbol{\lambda}$, and therefore they are not independent from each other. 
This implies that a prior on $\boldsymbol{\lambda}$ is induced by the PE prior choice, which may not be consistent with the astrophysically motivated prior $p_{\rm astro}(\boldsymbol{\lambda})$ (that will be described in Sec.~\ref{sec:Astrophysical_prior} below). 
Hence, we reweigh the posterior samples on source parameters of our merger events by the \emph{astrophysical} prior on $\boldsymbol{\theta}$, $p_{\rm astro}(\boldsymbol{\theta})$, which we introduce to be consistent with $p_{\rm astro}(\boldsymbol{\lambda})$. 
To obtain $p_{\rm astro}(\boldsymbol{\theta})$, we first draw samples $\{\boldsymbol{\lambda}_{i}\}$ from $p_{\rm astro}(\boldsymbol{\lambda})$, then we get the samples $\boldsymbol{\theta}_i={\cal F}(\boldsymbol{\lambda}_i)$ with forward modeling, and finally we estimate the function  $p_{\rm astro}(\boldsymbol{\theta})$ with kernel density estimation on these samples.

We reweigh the source posterior samples with probability weight $w=p_{\rm astro}(\boldsymbol{\theta}) / p_{\rm PE}(\boldsymbol{\theta})$. 
This procedure gives us a new set of binary-source posterior samples, consistent with the choice of astrophysical prior on $\boldsymbol{\lambda}$. 

\begin{figure*}
    \centering
    \includegraphics[width=0.49\textwidth]{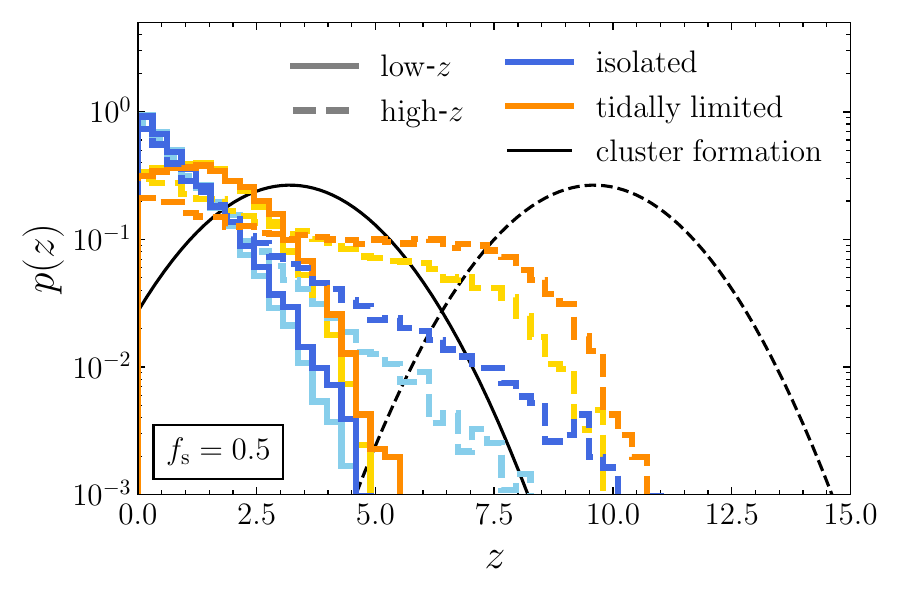}
    \includegraphics[width=0.49\textwidth]{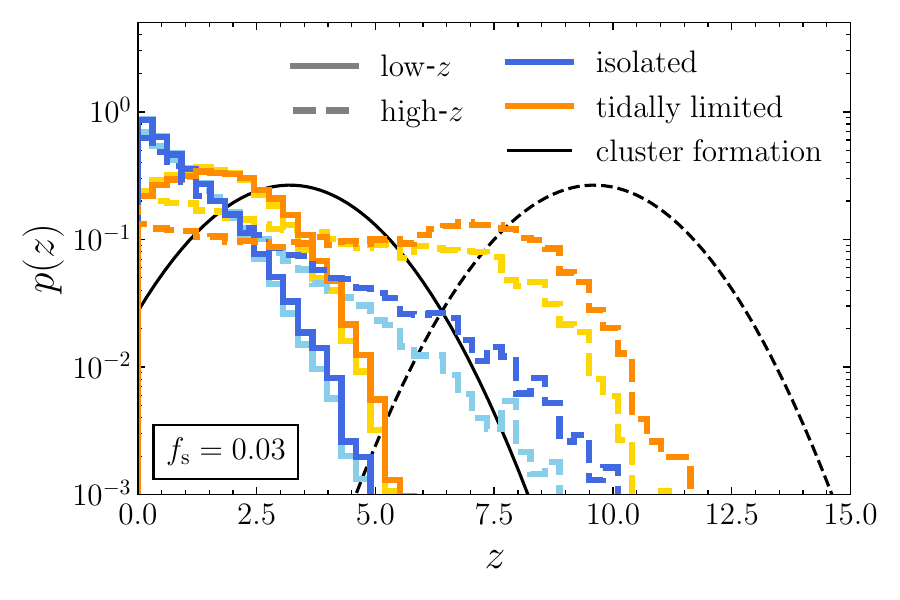}
    \caption{Probability density distribution of the merger redshifts of all IMBH binaries with masses between $100M_\odot$ and $1500M_\odot$ with $f_{\rm s}=0.5$ (left panel) and $f_{\rm s}=0.03$ (right panel). The cluster initial conditions have been sampled from the astrophysical prior (see Sec.~\ref{sec:Astrophysical_prior}). Two cluster formation histories have been assumed: a low-$z$ prior (black solid curve) and a high-$z$ prior (black dashed curve). The blue and orange curves correspond to the \emph{intrinsic} distributions in the isolated and tidally limited cases, while the cyan and yellow curves correspond to the \emph{detectable} distributions, respectively, computed by imposing a network SNR threshold of 12.}
    \label{Fig:redshift_astrophysical_prior}
\end{figure*}

\subsubsection{Astrophysical prior}
\label{sec:Astrophysical_prior}

We assume the priors on $M_{\rm cl,0}$, $r_{\rm h,0}$, and $z_{\rm cl,0}$ to be uncorrelated with each other. 
There is currently no observational support for a redshift evolution of the initial cluster mass function. The observed correlation between mass and radius in young massive clusters is weak, and therefore we ignore it here~\cite{Larsen:2003fp,2019ARA&A..57..227K}. 

As for the available growth time $\Delta t_{\rm g}$, we need to assume a specific astrophysical scenario to come up with a physically sensible prior distribution. 
Here, we consider the coalescence of clusters after they sink into the center of the host galaxy or a star-forming complex through dynamical friction as a physically motivating setting. 
Our method is insensitive to which cluster reached the center first. 
Therefore, we give equal probability to each of the two scenarios (the cluster of either the primary or the secondary IMBH reaches the center first), bearing in mind that the initial cluster mass $M_{\rm cl,0}$ and galactocentric radius $R_{\rm g,0}$ are coupled through the dynamical friction timescale, because $\tau_{\rm df}\propto R_{\rm g,0}^2/M_{\rm cl,0}$~\cite{1988gady.book.....B}. 
Notice that the cluster evolves as it inspirals into the center, hence in principle the dynamical friction timescale depends on the cluster properties in a nontrivial way~\cite{Antonini:2012sj}, but for simplicity we neglect these complications. 

We thus write the astrophysical prior as
\begin{widetext}
\begin{align}
    p_{\rm astro}(\boldsymbol{\lambda}) \propto {1\over2}\left[p(\Delta t_{\rm g,1} | M_{\rm cl,01}) + p(\Delta t_{\rm g,2} | M_{\rm cl,02}) \right] \times p(M_{\rm cl,01}) p(M_{\rm cl,02}) p(r_{\rm h,01}) p(r_{\rm h,02}) p(z_{\rm cl,01}) p(z_{\rm cl,02}).
    \label{Eq:astro_prior}
\end{align}
\end{widetext}
Notice that the factors $p(\Delta t_{\rm g,1}|\Delta t_{\rm g,2}, z_{\rm cl,01}, z_{\rm cl,02})$ and $p(\Delta t_{\rm g,2}|\Delta t_{\rm g,1}, z_{\rm cl,01}, z_{\rm cl,02})$, which arise as a consequence of Bayes' theorem, are both constants, and do not appear in the expression of the prior above. 
This is due to the time constraint of Eq.~\eqref{eq:time_constraint}.

To derive $p(\Delta t_{\rm g,i}|M_{\rm cl,0i})$ in Eq.~\eqref{Eq:astro_prior} we need a prior for the galactocentric radius. 
Taking a singular isothermal distribution for the volumetric number density of forming star clusters, we find that the cumulative number of clusters contained within radius $r$ is $N(<r)\propto r$. 
Thus, the probability density function of $R_{\rm g,0}$ is constant, as $dN(<r)/dr\propto \rm constant$. 
Using the dependence of the dynamical friction time and imposing that $\Delta t_{\rm g}=\tau_{\rm df}$, we arrive at
\begin{equation}
  p(\Delta t_{\rm g} | M_{\rm cl,0})\propto M_{\rm cl,0} / \sqrt{\Delta t_{\rm g}}\,.
\end{equation}

In the tidally limited model, since clusters may evaporate in a finite time before reaching the center, we make sure that the growth time $\Delta t_{\rm g}$ is smaller than the evaporation time $\tau_{\rm ev}$. 
Furthermore, clusters must reach the center after assembling the IMBH in their core, otherwise there will be no IMBH by the time the cluster coalesces with another cluster. This results in the constraint $\tau_{\rm BH}>\tau_{\rm df}$. 

Based on observations of young massive clusters in the local Universe, we assume a simple power-law distribution,
\begin{equation}
p(M_{\rm cl,0})\propto M_{\rm cl,0}^{-2}\,.
\end{equation}
While a Schechter mass function with a truncation mass scale at $\sim10^6M_\odot$ would also fit the data, there is no decisive statistical significance in favor of either model~\cite{2019ARA&A..57..227K,2014CQGra..31x4006K}. The choice of whether or not to include a truncation scale at $10^6M_\odot$ in the initial cluster mass function would not significantly impact our results, because the majority of the allowed parameter space volume lies below the truncation scale, unless the latter were much smaller than $10^6M_\odot$ on average (cf.~Figs.~\ref{Fig:Mcl0_rh0_MBH_} and~\ref{Fig:Mcl0_rh0_MBH_fs003}). 
We assume the initial cluster mass function to be universal and take its prior range to be $[10^4,10^7]M_\odot$. 
We choose a log-uniform prior on the half-mass radius in the range $[10^{-2}, 10]\,\rm pc$. 
The choice of cluster formation history relies on cosmological simulations of globular cluster formation~\cite{2019MNRAS.482.4528E,2022MNRAS.517.3144R}. 
Motivated by these studies, we model $p(z_{\rm cl,0})$ as a Gaussian distribution with mean $\overline{z}_{\rm cl,0}=3.2$ and standard deviation $\sigma_{z_{\rm cl,0}}=1.5$, as in Ref.~\cite{Mapelli:2021gyv}, defined in the prior domain $[0,15]$. 

In Fig.~\ref{Fig:redshift_astrophysical_prior} we compare the astrophysical prior on $z_{\rm cl,0}$ (black) with the resulting intrinsic distributions of $z_{\rm m}$ after applying our forward model in the isolated (blue) and tidally limited (orange) cases. The initial conditions for the cluster parameters have been drawn from the astrophysical prior, but the distributions are only shown for a population of IMBH binary mergers with masses in the range $[100, 1500]M_\odot$. 
We assume $f_{\rm s}=0.5$ in the left panel and $f_{\rm s}=0.03$ in the right panel, with the latter choice leading to slightly smaller time delays.
To highlight the importance of cluster formation history on the distribution of $z_{\rm m}$, we have chosen two different priors on $z_{\rm cl,0}$: a low-redshift one that peaks at $z=3.2$ (solid), and a high-redshift one which peaks at $z=9.6$ (dashed). 
Finally, we also show the redshift distributions for detectable binaries in the isolated and tidally limited cases (in cyan and yellow colors, respectively) assuming a network SNR threshold of 12. 
The orientation and sky position of each IMBH merger have been sampled isotropically.
While the selection effect only affects the redshift distribution at high redshifts, we may still be able to distinguish between a low-redshift and a high-redshift cluster formation scenario by measuring the highest redshift events in the tail of the distribution.

\subsection{Results}
\label{sec:Results}

In this subsection we present our results. We start with an example showing the full corner plot for binary D (Sec.~\ref{sec:Full_cluster_posterior_example}) and then we show the marginalized posteriors for the rest of the simulated events (Sec.~\ref{sec:Marginilized_cluster_posteriors_for_fs05}).

\subsubsection{Full cluster posterior example}
\label{sec:Full_cluster_posterior_example}

\begin{figure*}
    \centering
    \includegraphics[width=\textwidth]{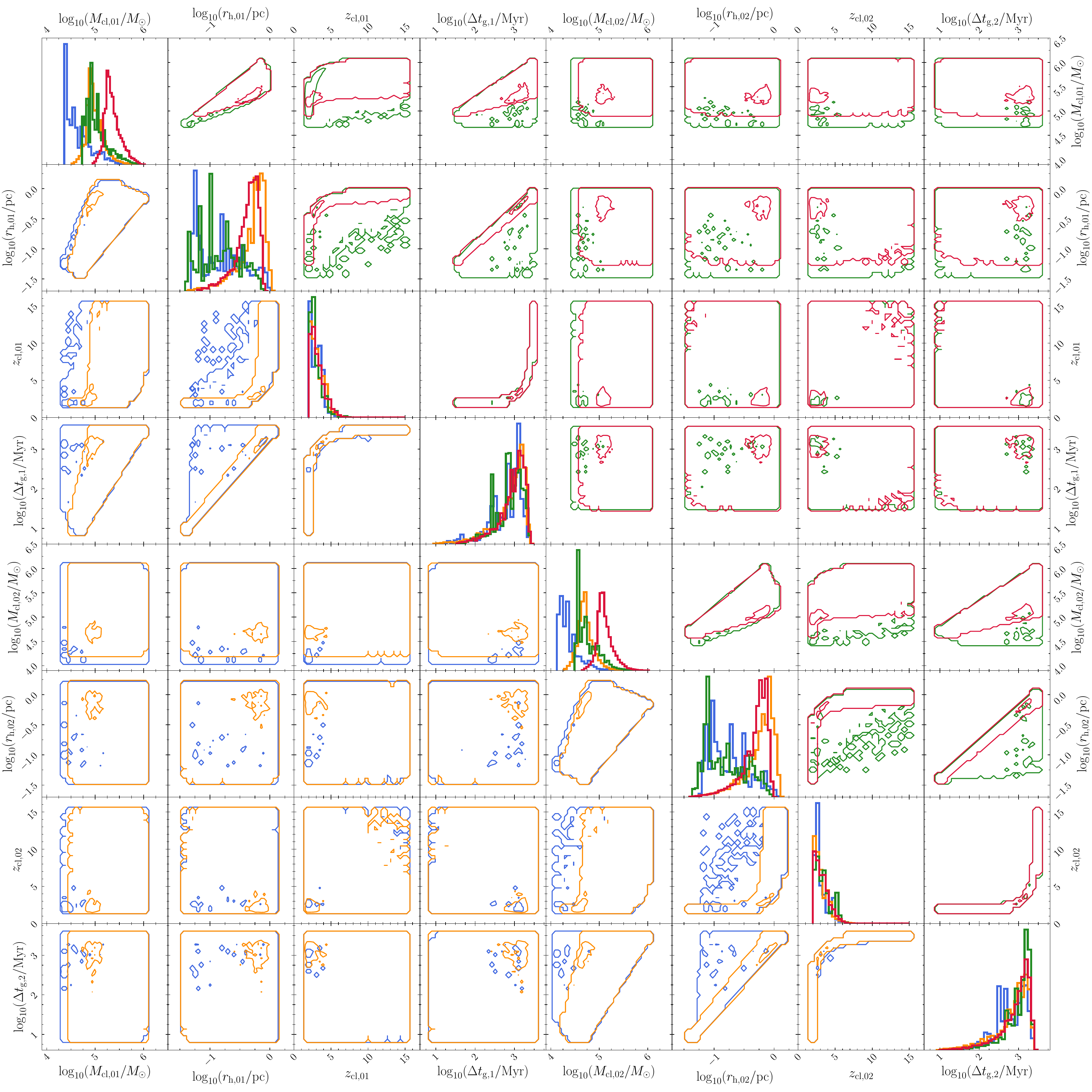}
    \caption{Cluster posterior distributions for the progenitors of the IMBH components of binary D (see Table~\ref{tab:binary_parameters}). The lower-left corner plot shows the isolated (blue) and tidally limited (orange) contours for $f_{\rm s}=0.5$. The upper-right corner plot shows the isolated (green) and tidally limited (red) contours for $f_{\rm s}=0.03$. The diagonal entries show the marginalized posteriors for all four cases. The two contours in each panel correspond to the $0.5\sigma$ (11.8\%) and $1.5\sigma$ (67.5\%) confidence levels.}
    \label{Fig:corner_plot_600_200_2}
\end{figure*}

In Fig.~\ref{Fig:corner_plot_600_200_2} we focus on IMBH binary D (cf. Table~\ref{tab:binary_parameters}) and we show the full posterior distributions of the parameters of the progenitor star clusters. 
While we show the full posterior results for only one binary as an example, the qualitative features described here apply to all binaries we examined.
The lower-left corner plot shows 2D posteriors on cluster parameters
assuming that a fraction $f_{\rm s}=0.5$ of each star's mass is consumed by the BH, while the upper-right corner plot assumes $f_{\rm s}=0.03$. The panels on the diagonal show the 1D marginalized distributions for both cases.
The two colors in the lower (upper) corner plot correspond to the two cluster evolutionary models considered in this study, i.e., the isolated case (in blue or green) and the tidally limited case (in orange or red), respectively.
In each panel, we show two contours, corresponding to $0.5\sigma$ (11.8\%) and $1.5\sigma$ (67.5\%) confidence levels. 

We will first focus on a specific case. Then we will discuss the difference between the isolated and tidally limited cases, and the effect of the accretion rate.

\noindent
{\bf \em The $f_{\rm s}=0.5$ isolated scenario.}
Let us first focus on the isolated scenario with $f_{\rm s}=0.5$ (blue contours in the lower-left corner plot).
While the IMBH masses and redshift of binary D are measured with an error of a few percent, we see that the structural properties of the star clusters (i.e., initial masses and half-mass radii) are constrained rather poorly.  The shape of the contours in the $M_{\rm cl,0}$--$r_{\rm h,0}$ plane is similar to the shape of the allowed parameter space (Fig.~\ref{Fig:Mcl0_rh0_MBH_}), with a preference for lighter cluster masses due to our prior choice.  Typical initial cluster masses $M_{\rm cl,0}\sim{\rm few}\times10^4M_\odot$ are favored for producing the IMBH masses of this particular binary ($m_1=600 M_\odot$ and $m_2=200 M_\odot$) through the runaway TDE scenario.  Note also that the posterior of $M_{\rm cl,02}$ has more support at a lower value than the posterior of $M_{\rm cl,01}$:
the inference suggests that the lighter BH formed in a lighter cluster.

The posteriors of $r_{\rm h,0}$ span a wide range of values, but there is a strong preference for compact clusters with radii on the order of $\sim0.1\rm pc$. This is related to the prior choice for the growth time $\Delta t_{\rm g}$ and initial cluster mass $M_{\rm cl,0}$. 
Since $p(\Delta t_{\rm g}|M_{\rm cl,0})\propto M_{\rm cl,0}/\sqrt{\Delta t_{\rm g}}$ (as a consequence of the time dependence of dynamical friction on the initial galactocentric radius and cluster mass), there is a preference for clusters that reach the center faster. Such systems have a smaller allowed growth timescale before merging with another cluster, and more compact systems evolve faster because the relaxation time depends more strongly on the radius than on the mass scale. 
Furthermore, the contour lines of constant BH mass on the cluster mass-radius plane have a positive slope, i.e., a preference for lighter clusters (as in the assumed prior) results in smaller values for $r_{\rm h,0}$. 
To summarize: the cluster mass controls the mass scale of the IMBH, while the half-mass radius controls the speed at which the asymptotic IMBH mass is approached. 

The posterior on the structural parameters is relatively broad mostly because of model degeneracies, rather than measurement uncertainties. 
As in Fig.~\ref{Fig:Mcl0_rh0_MBH_}, a contour of constant BH mass formed within a Hubble time is (to a good approximation) a line in the $M_{\rm cl,0}$--$r_{\rm h,0}$ parameter space, meaning that there are multiple clusters with different initial conditions whose IMBH mass asymptotes to the same value. 
The nontrivial structure in the cluster posteriors arises due to the time information. 
In particular, a specific cluster's radius strongly correlates with the growth time within the same system (the correlation is stronger for tidally limited clusters).
This is because the radius controls how fast clusters evolve: smaller-$r_{\rm h,0}$ clusters, that lose mass faster, need to merge with another cluster on a smaller timescale before they completely evaporate, hence the smaller growth times. 
On the other hand, there is no correlation between $\Delta t_{{\rm g},i}$ and $r_{{\rm h},0j}$ for $i\ne j$, as parameters belonging to different clusters are uncorrelated in our model.
The correlation between $\Delta t_{\rm g}$ and $z_{\rm cl,0}$ is also expected, since a larger growth time results in earlier star cluster formation times. 

In our framework, we find that the strongest constraints can be placed on the formation redshift of the clusters, in the sense that the marginalized posteriors on $z_{{\rm cl},0i}$ are relatively narrow compared to the width of the assumed prior on this parameter. 
We find that the marginalized posterior on $z_{{\rm cl},0i}$ is strongly peaked at redshifts close to the merger redshift of the IMBH binary, $z_{\rm m}$.
This is due to the choice of prior on $\Delta t_{\rm g}$. 
Clusters that form close to the center of the host galaxy preferably inspiral rapidly to the center, and they have to assemble the IMBH within the same time. 

\begin{figure*}
    \centering
    \includegraphics[width=\textwidth]{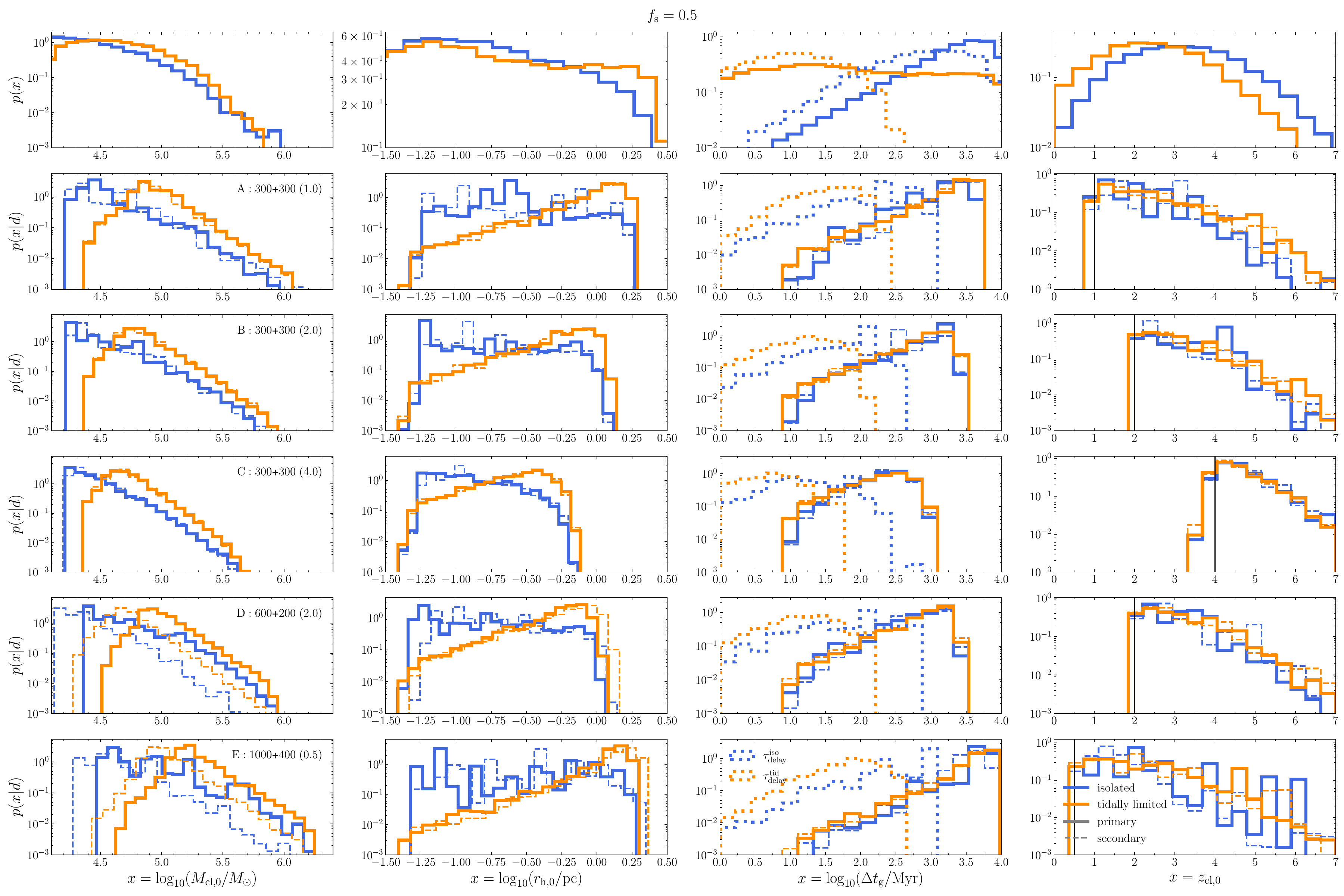}
    \caption{Marginalized cluster posterior distributions, assuming the low-redshift prior on $z_{\rm cl,0}$ and an accretion parameter $f_{\rm s}=0.5$. Different rows correspond to different binary systems (labeled A--E), while the first row shows the priors for each of the parameters. The numbers in the legend next to the model label indicate the binary component masses and (in parentheses) their redshift. Blue (orange) lines correspond to isolated (tidally limited) evolution; solid (dashed) lines show the posteriors associated with the progenitor cluster of the primary (secondary) IMBH, respectively. In order from the leftmost to the rightmost column, these are: initial cluster mass $M_{\rm cl,0}$, initial half-mass radius $r_{\rm h,0}$, growth time $\Delta t_{\rm g}$, and redshift of cluster formation $z_{\rm cl,0}$. The dotted distributions in the third column show $\tau_{\rm delay}$. The black vertical line in the rightmost panels is the merger redshift.}
    \label{Fig:cluster_posteriors_lowz_fs05}
\end{figure*}

\begin{figure*}
    \centering
    \includegraphics[width=\textwidth]{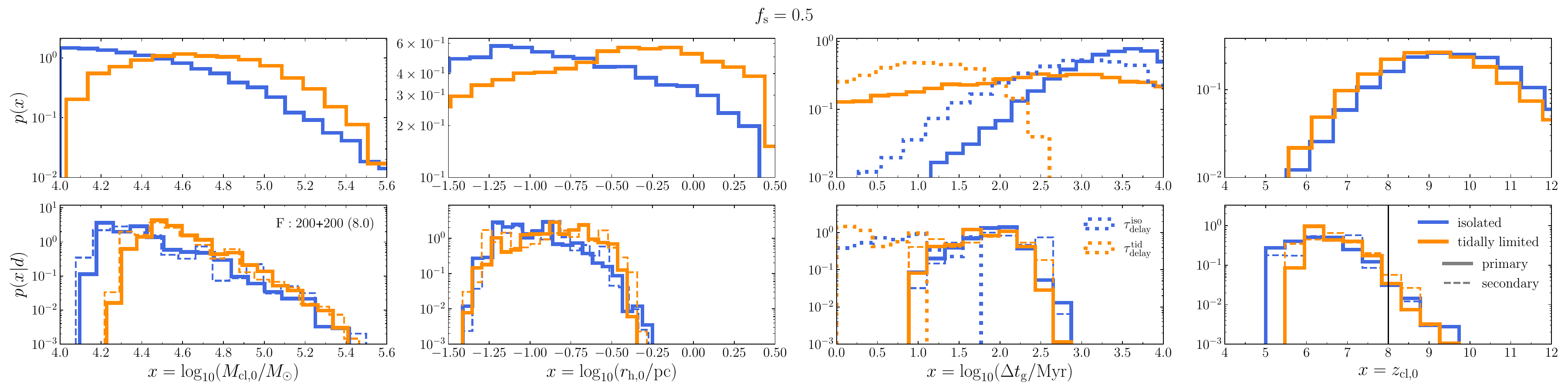}
    \caption{Same as Fig.~\ref{Fig:cluster_posteriors_lowz_fs05} but for binary F, and assuming the high-redshift prior on $z_{\rm cl,0}$.}
    \label{Fig:cluster_posterior_highz_fs05}
\end{figure*}

\noindent
{\bf \em Isolated vs. tidally limited clusters.}
In the isolated cluster evolution scenario (blue) we see a slight preference for lighter clusters relative to the tidally limited clusters (orange). This is because isolated clusters in our model have a smaller fraction of ejected stars with a velocity larger than the escape velocity, $\xi_{\rm e}$ (see Table~\ref{tab:cluster_evolutionary_scenarios}), and hence evaporate at a slower rate.

\noindent
{\bf \em Effect of the accretion rate.}
In the case of a lower accretion rate ($f_{\rm s}=0.03$, upper-right corner plot) we observe a preference for heavier star clusters relative to the case $f_{\rm s}=0.5$ discussed so far. This is because the BH growth rate is now smaller by a factor $\simeq 17$. 
Thus, since the asymptotic value of the BH mass
depends strongly on $M_{\rm cl,0}$, a heavier star cluster is required to produce the same BH mass. 
By looking at the diagonal in the corner plots of Fig.~\ref{Fig:corner_plot_600_200_2}, we observe that the peak in $r_{\rm h,0}$ for $f_{\rm s}=0.03$ also shifts to a higher value. 
This is due to a binning effect when projecting the posteriors onto the radius axis. Despite the higher density of points in the posterior for low values of $M_{\rm cl,0}$ and $r_{\rm h,0}$, there is a larger number of samples per logarithmic bin for higher values: see the panel in the first row and second column of Fig.~\ref{Fig:corner_plot_600_200_2}. This does not happen in the $f_{\rm s}=0.5$ case (second row, first column) because the parameter space ``volume'' in the high-density region is larger. This implies that the peak in $r_{\rm h,0}$ is at lower values.

There does not seem to be a significant variation in the distributions of the growth times, that have a typical value of $\sim1000\,\rm Myr$ with a long tail down to a few tens of $\rm Myr$.
For context, the Arches cluster, which is the densest young massive star cluster in the bulge of the Milky Way, with a galactocentric radius of $\sim30\,\rm pc$ and a total mass of $\sim10^4M_\odot$, will reach the center within a few hundred Myr (using an estimate based on the dynamical friction timescale of Eq.~(8.12) from~\cite{1988gady.book.....B}). 
Similarly, young super star clusters have been identified in the central regions of the starburst galaxy M82~\cite{McCrady:2003pe}, and their orbit is expected to decay through dynamical friction.

\subsubsection{Marginalized cluster posteriors for $f_{\rm s}=0.5$}
\label{sec:Marginilized_cluster_posteriors_for_fs05}

In this section, we present the marginalized one-dimensional posterior distributions of the cluster parameters and describe their features.
In Fig.~\ref{Fig:cluster_posteriors_lowz_fs05} we show the marginalized distributions for binaries A--E (rows two to six) computed assuming the low-redshift prior on $z_{\rm cl,0}$ for all of them (cf. Fig.~\ref{Fig:redshift_astrophysical_prior}). In Fig.~\ref{Fig:cluster_posteriors_lowz_fs05} we plot separately the results for binary F, computed assuming the high-redshift prior instead.
In both figures, the first row shows the marginalized prior for each parameter. 
In each of the panels we plot the posteriors of the two clusters hosting an IMBH, with solid (dashed) lines corresponding to the progenitor cluster of the primary (secondary) IMBH, respectively.
The colors correspond to the isolated (blue) and tidally limited (orange) evolutionary models. 
Although $\tau_{\rm delay}$ is not an independent parameter of our model, we show its distribution with dotted histograms (along with the $\Delta t_{\rm g}$ distributions) in the third column. 
All of these plots refer to an accretion parameter $f_{\rm s}=0.5$. To avoid clutter, the corresponding marginalized posteriors for the case $f_{\rm s}=0.03$ are shown in Appendix~\ref{App:Margninalized_cluster_posteriors_for_fs003}.

As a general trend, we find that higher redshift systems require denser clusters. 
This can be seen by comparing binaries A, B, and C, which have the same component masses but different merger redshifts:
binary A merges at $z_{\rm m}=1$ (when the age of the Universe is $t\simeq5.8\,\rm Gyr$), while binary C merges at $z_{\rm m}=4$ ($t\simeq1.5\,\rm Gyr$). 
There is less time available to assemble the IMBHs in binary C, ensuring that they pair and eventually merge, hence more compact stellar cluster progenitors are required.
In particular, the posteriors on the radii inferred for systems A to C assuming tidally limited evolution peak at $r_{\rm h,0} \sim1.3\,\rm pc$, $\sim0.7\,\rm pc$, and $\sim0.4\,\rm pc$, respectively.
In addition, tidally limited clusters evaporate at a higher rate, and thus require heavier and wider systems that survive for longer. As a consequence, the priors $p(M_{\rm cl,0})$ and $p(r_{\rm h,0})$ shown in the top row of Fig.~\ref{Fig:cluster_posteriors_lowz_fs05} have a preference for high values.

As the redshift is lowered at fixed IMBH masses (i.e., moving ``upward'' in the plot from binary C to binary A), the accessible parameter space in the $M_{\rm cl,0}$--$r_{\rm h,0}$ plane grows to larger radii: compare the left and right columns of Fig.~\ref{Fig:Mcl0_rh0_MBH_}. This is why the posteriors on $r_{\rm h,0}$ gains more support at larger values. 
Since the initial cluster mass and radius are correlated, the posteriors on $M_{\rm cl,0}$ also gain support at higher masses. 
Notice that for these systems there is no statistical difference between the solid and dashed curves (corresponding to the primary and secondary IMBH): the IMBHs in these binaries have the same mass,
and due to symmetry there should no be difference in the induced distributions for their progenitor clusters (although the two IMBHs might have been assembled in clusters with different initial conditions due to model degeneracy). 
The peak that appears in $p(r_{\rm h,0}|d)$ occurs due to the flattening of the $M_{\rm BH}$ contours as they approach the $\tau_{\rm BH}=\tau_{\rm Hub}-t_{\rm cl,0}$ boundary (see Fig.~\ref{Fig:Mcl0_rh0_MBH_}), which causes a pile-up feature upon projection of the posterior samples onto the one-dimensional probability distribution corresponding to $r_{\rm h,0}$.

Binary E is interesting as it is the heaviest and closest event analyzed, merging at $z_{\rm m}=0.5$. 
This binary has the largest available lookback time and requires on average heavier clusters, especially to assemble the high-mass ($1000M_\odot$) primary IMBH. 
The cluster mass distributions for the secondary are shifted to slightly lower values, since the lower-mass ($400M_\odot$) secondary IMBH is preferentially assembled in lighter clusters.

As mentioned in the previous section, $r_{\rm h,0}$ is strongly correlated with $\Delta t_{\rm g}$. 
Thus, features in the radius posteriors (such as bimodality) are usually observed also in the posteriors of the growth time. 
The correlations are primarily driven by the prior on $\Delta t_{\rm g}$, which prefers small values relative to $\tau_{\rm delay}$ (see the third column of Fig.~\ref{Fig:cluster_posteriors_lowz_fs05}):
if the cluster has to quickly merge with another cluster, then it must be very compact and evolve rapidly in order to grow the IMBH within the limited time window. 

The total delay time, from cluster formation to binary merger, is dominated by the IMBH-IMBH formation and evolution timescale after the two clusters coalesce, i.e., by the $\tau_{\rm delay}$ timescale (dashed histograms in the third column of Fig.~\ref{Fig:cluster_posteriors_lowz_fs05}). 
Recall that in our astrophysical scenario of interest IMBHs grow rapidly in dense star clusters, which then quickly sink into the central regions of the star-forming complex by dynamical friction, where they form an IMBH binary that hardens and eventually merges. 
The first growth process can be rapid (within hundreds of Myr), while the phase of binary formation and evolution can take up to billions of years. 
For IMBH binaries that merge at high redshifts, however, the inferred delay times are required to be low (on the order of hundreds of Myr) because of the limited lookback time available.
Tidally limited clusters also have a preference for smaller delay times, because the cluster resulting from the merger of the two progenitor clusters is generally more compact and has a smaller relaxation timescale. 

The $z_{\rm cl,0}$ distributions in the fourth column of Fig.~\ref{Fig:cluster_posteriors_lowz_fs05} have the smallest variance when compared with the prior $p(z_{\rm cl,0})$. 
In this sense, the redshift at cluster formation is the best-constrained parameter. 
The constraining power becomes less significant for binary F (see Fig.~\ref{Fig:cluster_posterior_highz_fs05}), whose binary source parameters are worst measured due to its high redshift. 
Moreover, at those high redshifts, due to the nonlinear relation between redshift and cosmic time, there is less available physical time,
and the total delay time from cluster formation to the binary merger cannot be more than $\simeq400\,\rm Myr$ (which corresponds to the difference between the ages at $z=15$ and $z=8$). 
Although there is a higher probability for the clusters to have formed closer to the IMBH binary merger, there is a long tail in the cluster redshift distributions, which originates from the high-end tail in $\Delta t_{\rm g}$. 
Notice that the redshift distributions sometimes have support below the merger redshift due to the broad posterior on the merger redshift $z_{\rm m}$. This feature is most noticeable in the high-redshift events, whose parameters are harder to measure. 
Finally, the $z_{\rm cl,0}$ distributions are not very different even when the merging IMBHs have different masses. 
We have found that differences can be observed only when the mass ratio $q=m_1/m_2 \geq 10$, and that the distribution of $z_{\rm cl,0}$ weakly depends on the IMBH mass ratio.

\section{Limitations of the model}
\label{sec:Limitations_of_the_model}

In this section, we discuss the limitations of our model. 
In Sec.~\ref{sec:Alternative_IMBH_growth_channels} we present alternative formation channels that could form IMBHs. 
In Sec.~\ref{sec:Isothermal_assumption} we revisit the assumption of isothermal conditions in the cluster beyond the core radius.
In Sec.~\ref{sec:Validity_of_the_full-loss-cone_model} we discuss the validity of the TDE rate estimate of Eq.~\eqref{Eq:consumption_rate}. 
In Sec.~\ref{Sec:Complex_star_cluster_evolution}, we discuss the limitations of our assumption of a monolithic star cluster evolution composed of a single mass component.
Finally, in Sec.~\ref{sec:Comparison_with_N-body_simulations} we compare the predictions on the IMBH mass from our astrophysical model with the $N$-body simulations of Ref.~\cite{Rizzuto:2022fdp}. 

\subsection{Alternative IMBH growth channels}
\label{sec:Alternative_IMBH_growth_channels}

In this work, we have considered the formation of IMBHs through runaway tidal encounters. 
However, hundred-$M_\odot$ BHs can form in other ways. 
In this subsection, we review some of these alternative formation pathways, such as remnants of Pop III stars (Sec.~\ref{sec:Population_III_remnants}), runaway stellar collisions (Sec.~\ref{sec:Runaway_stellar_collisions}), repeated BH mergers (Sec.~\ref{sec:Repeated_BH_mergers}), and gas accretion (Sec.~\ref{sec:Gas_accretion}). 

\subsubsection{Population III remnants}
\label{sec:Population_III_remnants}

The first stars (known as Pop III stars) are believed to have weaker winds and their collapse will lead to heavier remnants~\cite{Madau:2001sc}. 
If formed in pairs, such stars may form BH binaries with both components masses above $\sim150M_\odot$, as suggested by population synthesis codes~\cite{Mangiagli:2019sxg,Hijikawa:2021hrf}. 
While the stellar winds of very massive stars are uncertain, and the details of the formation of Pop III stars are debated, the collapse of Pop III stars could contaminate the lower end of the IMBH mass spectrum, producing IMBH binary mergers at higher redshifts~\cite{Ng:2020qpk}. 
If the cluster formation rate peaks at $z\sim4$, then the population of IMBH binary mergers from gravitational runaways studied in this paper could be distinguished from massive Pop III binaries, whose merger rate would peak at around $z\sim10$.
In addition, the collapsar model presented in Ref.~\cite{Siegel:2021ptt} could also form IMBHs beyond $130M_\odot$. 

\subsubsection{Runaway stellar collisions}
\label{sec:Runaway_stellar_collisions}

Stars with masses $\gtrsim 200M_\odot$ formed through runaway stellar collisions may also lead to the formation of IMBHs~\cite{PortegiesZwart:1998nb,PortegiesZwart:2004ggg,Gurkan:2005xz,Prieto:2024pkt,doi:10.1126/science.adi4211}.  However, the mass growth rate of the runaway star may be too strong for a very massive star to even form in a metal-rich environment~\cite{2009A&A...497..255G,Mapelli:2016vca}, so this formation scenario may only be relevant in low-metallicity systems.

A simple analytic model for the production of IMBHs through this channel, where the mass of the IMBH progenitor is given as a function of the cluster's initial mass, was developed in Ref.~\cite{PortegiesZwart:2002iks} and verified through $N$-body simulations. 
Assuming that the very massive star assembled via this mechanism collapses directly into an IMBH, and neglecting any stellar wind effects that would reduce the mass of the star by the end of its life, Eq.~(16) from~\cite{PortegiesZwart:2002iks} finds the simple relation $M_{\rm IMBH}\simeq m_{\rm seed} + 0.008M_{\rm cl,0}$. Here $m_{\rm seed}$ is the zero-age main-sequence seed mass of the heaviest star that initially forms in the cluster, which may be $\sim150M_\odot$~\cite{Figer:2005gr}. 
Based on this equation, the mass scale of star clusters that contribute to the formation of hundred-$M_\odot$ IMBHs is similar to the mass scale of the runaway TDE channel examined in this paper for $f_{\rm s}=0.5$. 
Nevertheless, the stellar collision channel requires sufficiently small initial relaxation times (hence, very compact stellar systems), so that the process can kick-start before the heaviest stars form a BH subsystem that can prevent such a stellar runaway. 
The formation of the IMBH through stellar collisions is very rapid (within only a few Myr), and based on the classification of Ref.~\cite{2015MNRAS.454.3150G} it would correspond to a fast growth scenario. 

The IMBHs formed through this channel may be ejected from the cluster in its later evolutionary phases. 
Reference~\cite{Prieto:2022uot} found a very low retention of IMBHs with a mass in the hundred solar masses in globular clusters: in their simulations, all IMBHs formed through runaway stellar collisions are ejected within $500\,\rm Myr$. 
Since IMBHs form very rapidly (within only a few Myr before the formation of the BH subsystem), they typically interact with lighter BHs, possibly leading to intermediate-mass ratio inspirals~\cite{Mandel:2007hi}. 
The interaction of the IMBH with other stellar-mass BHs can induce a moderate kick of the order of hundreds of $\rm km\, s^{-1}$, that may still be larger than the escape velocity. 
Similar kicks can also result from the relativistic recoil of the remnant of an IMBH-BH merger. 
The IMBH may be retained in the system if the escape velocity is high enough (for example, in nuclear star clusters~\cite{Neumayer:2020gno}). 
On the other hand, Ref.~\cite{Rizzuto:2022fdp} found efficient retention of IMBHs that grow through runaway TDEs, and their simulations did not eject any IMBH. 
This is because the Newtonian kick imparted to the IMBH is not large enough to eject it as a consequence of momentum conservation. After all, the mass ratio is very asymmetric (typically, $q>100$) during IMBH-star-star interactions. 
Nevertheless, interactions of the IMBH with binary stars are more likely to produce hyper-velocity stars ejected from the cluster through the Hills mechanism~\cite{1988Natur.331..687H}. 

\subsubsection{Repeated BH mergers}
\label{sec:Repeated_BH_mergers}

As discussed in Sec.~\ref{sec:Dependence_on_cluster's_initial_condition}, star clusters with a sufficiently high escape velocity undergo runaway BH mergers (see also~\cite{Atallah:2022toy,Zevin:2022bfa,Kritos:2022non}). 
The simulations shown in the right panel of Fig.~\ref{Fig:vesc_mbhmax__m1_q} suggest that the formation of two IMBHs with masses between $200M_\odot$ and $1000M_\odot$ within the same stellar environment, while possible, is unlikely (see also, e.g., Fig.~1 from Ref.~\cite{Fragione:2021nhb} for an independent study supporting this conclusion). 

In the cluster merger scenario, two IMBHs in the mass range considered in this paper can still form through the repeated BH merger channel in very heavy and compact stellar clusters. 
It is hard to find unambiguous observational signatures that could tell the two possibilities apart. Spin measurements are unlikely to help, as the spin of a BH growing from minor incoherent accretion episodes always asymptotes to zero~\cite{Hughes:2002ei,King:2008au,Berti:2008af} (see also Appendix~\ref{App:mass_spin_evolution}).
However, one could argue that the heavier star clusters required for runaway BH mergers are much rarer than lighter systems, and the corresponding merger rates would be  lower. For cluster with escape velocity below $\sim 200\,\rm Km\,s^{-1}$, the left panel of Figure~\ref{Fig:vesc_mbhmax__m1_q} suggests that the TDE channel should dominate, in the sense that IMBHs are unlikely to form through repeated BH mergers for those clusters.

\subsubsection{Gas accretion}
\label{sec:Gas_accretion}

Observational data imply that, despite being dry stellar systems, globular clusters contain some gas in their centers, resulting from stellar winds of low-mass stars and ionized by ambient white dwarfs. 
For instance, Ref.~\cite{Abbate:2018pdf} inferred a density of gas particles $n_{\rm e}=0.23\,\rm cm^{-3}$ using the dissipative effects that the gas would have on the emission of millisecond pulsars in 47 Tucanae. 

A stellar-mass BH of $10M_\odot$ accreting at the Bondi rate in the core of 47 Tucanae (assuming a spherical flow) would not grow to hundreds of solar masses within a Hubble time.
Therefore, the gas contained in evolved clusters does not reach densities suitable to produce an IMBH through accretion.
Nevertheless, if accreting continuously at the Eddington rate, a $10M_\odot$ BH would grow into the hundreds of $M_\odot$ within $\sim200\,\rm Myr$, assuming a Salpeter timescale of $45\,\rm Myr$.
This channel (in which IMBHs grow via gas accretion in gas-rich nuclear star clusters) was considered in Ref.~\cite{Natarajan:2020avl}, and it may contaminate the formation channel examined in this work. 
Spin measurements could be useful to distinguish the two scenarios: the spin of the IMBH would be driven to near-extremal values if accretion is coherent (rather than episodic), i.e., it would be much larger than the typical spin magnitudes of $0.1$ found in the channel proposed in this work (cf. Appendix~\ref{App:mass_spin_evolution}).

\subsection{Isothermal assumption}
\label{sec:Isothermal_assumption}

The assumption of isothermal conditions in $r_{\rm a}\lesssim r\lesssim r_{\rm h}$ is valid as long as $r_{\rm a}$ is smaller than the core radius $r_{\rm c}$. Otherwise, the core velocity dispersion within the cusp region is enhanced by the presence of the central BH.
Based on numerical simulations and an energy balance between the flux through $r_{\rm h}$ and $r_{\rm c}$ in systems with a central BH, Ref.~\cite{Heggie:2006se} derives that $r_{\rm c}/r_{\rm h}\simeq4.3(M_{\rm BH}/M_{\rm cl})^{3/4}$.
Thus, $r_{\rm c}/r_{\rm a}\simeq0.6(M_{\rm cl}/M_{\rm BH})^{1/4}$. 
Since we are interested in IMBHs with a mass no larger than $2000M_\odot$ and clusters with $r_{\rm h,0}>0.1\,\rm pc$, we estimate that the ratio $r_{\rm c}/r_{\rm a}$ does not fall short of unity. At the same time, it has a typical value of $\approx1.9$ in the parameter space region of interest for this work.
Therefore, the enhancement in $\sigma$ is not an issue, because we deal with relatively light IMBHs.

Within the influence radius of the central growing IMBH, we have assumed Bahcall-Wolf conditions with density power law index $\gamma=7/4$, as verified in $N$-body simulations. 
Nevertheless, according to Ref.~\cite{Alexander:2008tq} $\gamma$ can be as low as $1.4$ in the presence of other compact objects instead, and the consumption rate in that case would drop by $\simeq64\%$.
This would lead to a slower growth rate and cause a systematic effect in the measurement of the cluster properties. 
Since the asymptotic value of the IMBH mass is more strongly linked to the initial cluster mass, a larger value of $\gamma$ would result in an underestimation of $M_{\rm cl,0}$. 

\subsection{Validity of the full loss-cone model}
\label{sec:Validity_of_the_full-loss-cone_model}

Our stellar consumption rate formula [Eq.~\eqref{Eq:consumption_rate}], while conservative, still assumes an instantaneous repopulation of loss-cone orbits with energy $\sim m_\star\sigma^2$ (which corresponds to an orbit with semimajor axis equal to $r_{\rm a}$). 
Nevertheless, according to Fig.~6.5(b) in~\cite{Merritt2013}, the loss cone is efficiently filled by two-body relaxation for small BH masses $\lesssim10^4M_\odot$, because the $q$ parameter (which measures the degree to which the loss cone is filled by gravitational relaxation) is greater than unity for most stars inside $r_{\rm a}$ with energies $>m_\star \sigma^2$.
Essentially, the relaxation time in the core is small enough that star-star encounters efficiently cause energy and angular momentum diffusion, and provide a steady flux of stars into the loss cone~\cite{1976MNRAS.176..633F,1977ApJ...211..244L}.
Thus, we expect our full loss-cone model to approximate $\Gamma_{\rm C}$ reasonably well.

\subsection{Complex star cluster evolution}
\label{Sec:Complex_star_cluster_evolution}

In our model, we consider the monolithic formation of star clusters and neglect the effect of continual star formation. Nuclear star clusters, for example, are believed to have more complex star formation histories~\cite{Neumayer:2020gno}, and subsequent star formation could be an additional form of energy. In that case, the evolution of the cluster's mass and radius would be different from the one predicted in this work, and our formalism could lead to systematic bias in the inference of the cluster's initial conditions.

The relaxation time in our model assumes a single-mass cluster. To include the effect of a more complex mass function, we may divide the relaxation time by a multimass factor $\psi$ leading to a different mass-loss rate~\cite{2023MNRAS.522.5340G}. While this may be $\sim100$ for a Kroupa initial mass function, it rapidly decreases to a few within $10\,\rm Myr$, as massive stars evolve into remnants. In the case of a two-mass model containing stars and BHs, $\psi\simeq5$~\cite{Antonini:2019ulv}. This factor approaches unity as BHs are ejected, and the growth of the IMBH via repeated TDEs initiates because, by that time, the cluster mass spectrum is narrow. Additionally, our assumed timescale for $\tau_{\rm BH}$ follows Ref.~\cite{Breen:2013vla}, which is calculated assuming a two-mass model and accounts for the effect of BHs on the relaxation of the whole cluster. The inclusion of a time-evolving $\psi$ factor in our model would lead to a higher expansion rate for the clusters than what is predicted by Eq.~\eqref{Eq:drhdt}.

Furthermore, we have assumed that the BH subsystem evaporates on a timescale smaller than the evaporation rate of stars. However, $N$-body simulations carried out in Ref.~\cite{Banerjee:2011fa} suggest that if the BH natal kick is small enough for clusters to retain the majority of BHs, then those clusters within the inner regions of the Galaxy ($\lesssim5\,\rm kpc$) can evolve into a dark star cluster state during which the majority of the stars are ejected, while the remaining few are bound compactly to the dominant BH population.
The formation of IMBHs through repeated TDEs requires a small number of BHs embedded in a cluster dominated by stars, which cannot be realized in a dark star cluster.
A dark star cluster state could be avoided if, for example, BHs tend to receive a large natal kick, so most BHs are ejected at birth.

\subsection{Comparison with $N$-body simulations}
\label{sec:Comparison_with_N-body_simulations}

It is useful to compare our BH growth model through runaway tidal encounters with the $N$-body simulations performed in~\cite{Rizzuto:2022fdp}.
Since the authors of Ref.~\cite{Rizzuto:2022fdp} assume a single BH mass seed in the cluster, the system does not go through a phase of BH domination, and thus the growth of the BH starts promptly. 
Since we ignore the intermediate phase of a BH subsystem, the half-mass radius is smaller when the BH starts to grow via the consumption of stars.
As a consequence, the TDE rate will be higher, resulting in heavier BHs than estimated in the previous sections of this paper. 
To have as close a comparison as possible, in this subsection, we neglect the time delays due to core collapse and BH subsystem evaporation, i.e., we set $\tau_{\rm cc}=\tau_{\rm BH}=0$.

Moreover, Reference~\cite{Rizzuto:2022fdp} assumes a Kroupa initial mass function in the range $[0.08, 2.00]M_\odot$, which corresponds to a multimass relaxation factor of $\psi\simeq3$.
The initial number of stars in all systems is set to $N_{\star,0}=256000$, while $r_{\rm h,0}$ and $M_{\rm BH,0}$ are varied, for a total of five cluster models. For instance, their model $\rm R06M300$ has $r_{\rm h,0}=0.6\,\rm pc$ and $M_{\rm BH,0}=300M_\odot$ (see the second and third columns of Table~2 in~\cite{Rizzuto:2022fdp} for the values of the initial parameters).

\begin{figure}
    \centering
    \includegraphics[width=0.49\textwidth]{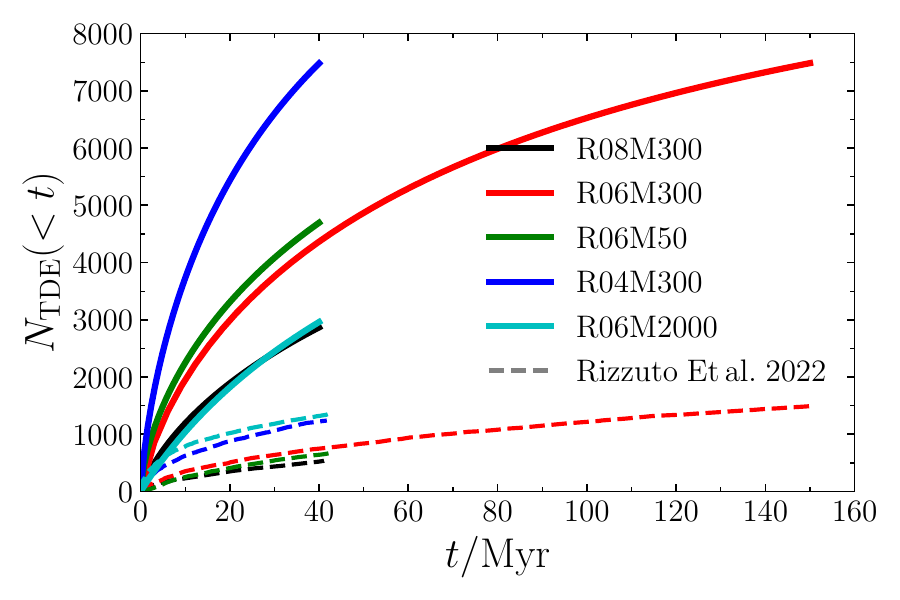}
    \caption{Predictions for the cumulative number of TDEs as a function of time with our BH growth model (thick solid lines) for the five clusters considered in Ref.~\cite{Rizzuto:2022fdp} whose simulated results are shown with thin dashed lines (compare their Fig.~13).}
    \label{Fig:predictions_of_Nbody_models}
\end{figure}

In Fig.~\ref{Fig:predictions_of_Nbody_models} we plot the number of TDEs as a function of time for those five cluster models.
In particular, we define $N_{\rm TDE}(<t)=[M_{\rm BH}(t) - M_{\rm BH,0}]/(f_{\rm s}m_\star)$ as the cumulative number of stars consumed within time $t$.
The value of $N_{\rm TDE}$ predicted with our BH growth model, and shown in Fig.~\ref{Fig:predictions_of_Nbody_models}, be compared with Fig.~13 of Ref.~\cite{Rizzuto:2022fdp}.
Despite all the differences in modeling, we find that our full loss-cone BH growth model agrees with~\cite{Rizzuto:2022fdp} within a factor of $\sim5$, generally leading to a slight overestimation of the BH mass.

\section{Conclusions}
\label{sec:Conclusions}

In this paper, we have studied the possibility of inferring the properties of star clusters by measuring the parameters of merging IMBHs with XG GW observatories, such as ET and CE. 
In our astrophysical scenario, the IMBHs are formed through runaway consumption of stars by a central growing BH in each cluster's core. The formation of IMBH binaries is assumed to follow the coalescence of the star clusters, each of which has formed a single IMBH. Our simple astrophysical model for cluster expansion and BH growth via runaway tidal encounters relies on the well-established balanced evolution. We have derived a simple, closed-form solution for the BH mass as a function of time, and we have verified its accuracy by numerical integration. The inference of the cluster properties is then possible because the masses of the IMBHs and the redshift of the merging binary are correlated with the initial conditions of the clusters (in particular, their initial masses and half-mass radii). We have discussed the limitations of our study and we have presented a list of alternative astrophysical channels through which such IMBH binaries may form. 

We have focused on IMBHs with component masses in the range $\sim100M_\odot$--$1000M_\odot$, because these will be detectable with XG ground-based GW observatories.
In principle, IMBH-BH mergers within clusters could also yield valuable information about their formation environment. However, we preferred to focus on comparable-mass binaries because current gravitational waveform models for systems with large mass ratios are unreliable. 
Binaries with even heavier IMBH components ($\gtrsim 1000M_\odot$) will require detectors with broadband sensitivity at decihertz frequencies~\cite{Sedda:2019uro}. There are many proposals to cover this GW frequency window, including space-based detectors such as ALIA~\cite{Bender:2013nsa,Baker:2019pnp} and DECIGO~\cite{Sato:2017dkf,Kawamura:2020pcg}; lunar experiments such as the Gravitational-wave Lunar Observatory for Cosmology (GLOC)~\cite{Jani:2020gnz}, the Lunar Gravitational Wave Antenna (LGWA)~\cite{Ajith:2024mie}, and the Lunar Seismic and Gravitational Antenna (LSGA)~\cite{Branchesi:2023sjl}; or experiments based on atom interferometry, such as AEDGE~\cite{AEDGE:2019nxb}, AION~\cite{Badurina:2019hst},  MAGIS~\cite{2019BAAS...51c.453H}, MIGA~\cite{2018NatSR...814064C}, ELGAR~\cite{Canuel:2019abg}, and ZAIGA~\cite{2020IJMPD..2940005Z}.

The runaway consumption of stars from IMBHs growing in the centers of star clusters has interesting observational consequences beyond GW astronomy~\cite{Komossa:2004vr}. 
The absence of an identified off-nuclear optical TDE counterpart out to a luminosity distance of $120\,\rm Mpc$ over 68 months of observations with the Zwicky Transient Facility constrains the number of TDEs per compact stellar system to be $<10^{-7}\,\rm yr^{-1}$~\cite{2024MNRAS.tmp..979P}, two orders of magnitude smaller than the observed nuclear TDE rate. 
From Eq.~\eqref{eq:dMBHdt} we can find a simple expression for the mean TDE rate per cluster:
\begin{align}
    \overline{\Gamma}_{\rm TDE}(<t) &= {M_{\rm BH}(t)\over f_{\rm s}m_\star t}\nonumber\\&\simeq3\times10^{-7}\,{\rm yr}^{-1} {M_{\rm BH}(t)\over 500M_\odot} {0.5\over f_{\rm s}}{0.3M_\odot\over m_\star}{10\,{\rm Gyr}\over t}\,.
    \label{Eq:TDE_rate}
\end{align}
This estimate, which is of the same order as the current observational constraint, is an upper bound, because full loss-cone theory overestimates the rate (see Sec.~\ref{sec:Comparison_with_N-body_simulations}). 
Moreover, most evolved globular clusters hosting an IMBH in their centers will likely be in their saturated regime, and the TDE rate will be much lower than the average value of Eq.~\eqref{Eq:TDE_rate}, which is dominated by the short runaway growth phase. 
The detection of one off-nuclear TDE by an IMBH candidate at a distance of $247\,\rm Mpc$ is the only such x-ray flare that is associated with a massive star cluster~\cite{2018NatAs...2..656L}.
In addition, the TDEs considered in this work focus on IMBHs with a mass so small 
that the IMBH is typically embedded within the stellar body~\cite{Gezari:2021bmb}; thus, the electromagnetic signatures of such TDEs are even less certain.
It is conceivable that, in the near future, the multimessenger combination of TDE signatures and GW observations of IMBH mergers will give us valuable information about the runaway growth of IMBHs through the repeated consumption of stars in star clusters.

A limitation of our proposal is that the initial mass and half-mass radius of the clusters are not strongly constrained due to model degeneracy, i.e., their posterior distributions are broad over the assumed prior domain. However, the redshifts of cluster formation are more narrowly constrained.
Therefore, given a population of IMBH binaries merging at different redshifts, one may infer the posteriors for cluster formation and reconstruct the cluster formation history by estimating their (model-dependent) delay time distribution. These population studies are an interesting topic for future research.

According to numerical simulations, the formation of globular clusters likely peaks at a redshift of around 3 to 4, i.e., before cosmic noon~\cite{2019MNRAS.482.4528E,2022MNRAS.517.3144R}. 
On the observational side, the Hubble Space Telescope and the James Webb Space Telescope (JWST) have an angular resolution limit that is not better than $\sim0.01''$ in their most sensitive frequency range. 
This corresponds to proper source-frame separations of order a few hundred parsecs at cosmological distances ($z>1$). 
Thus, direct electromagnetic probes of distant star clusters with radii below $10\,\rm pc$, which cannot be resolved individually, are currently challenging. 
Despite these limitations, a few strongly lensed young star clusters at $z>1$ have been identified with JWST~\cite{2024arXiv240103224A,2023ApJ...945...53V,2024arXiv240410770B}. 
These observations show evidence for the formation of young massive clusters at sub-pc scales that could grow IMBHs through runaway tidal encounters. 
The inference scheme proposed in this paper suggests that the IMBH merger rate, as well as the masses and redshift of the observed IMBH merger events, can be used to trace the formation history of star clusters. 
In future work, we will study the constraints that XG GW observatories can place on the history of cluster formation. 

A {\sc Python} implementation of our code for reproducibility of the results of this work is publicly available on {\sc GitHub} at the~\cite{githublink}.

\acknowledgments

We thank Itai Linial, Brian Metzger, Francesco Iacovelli, Digvijay Wadekar, Mark Cheung, and Yossef Zenati for useful discussions.
K.K. is supported by the Onassis Foundation - Scholarship ID: F ZT 041-1/2023-2024. 
K.K., L.R., and E.B. are supported by NSF Grants No. AST-2307146, No. PHY-2207502, No. PHY-090003, and No. PHY-20043, by NASA Grants No. 20-LPS20-0011 and No. 21-ATP21-0010, by the John Templeton Foundation Grant 62840, by the Simons Foundation, and by the Italian Ministry of Foreign Affairs and International Cooperation Grant No.~PGR01167.
K.K.Y.N. is supported by a Miller Fellowship and by the Croucher Foundation.
This work was carried out at the Advanced Research Computing at Hopkins (ARCH) core facility~\cite{rockfishlink}, which is supported by the NSF Grant No.~OAC-1920103.

\appendix

\section{Tidal capture radius}
\label{App:Tidal_capture_radius}

Stars on parabolic orbits that approach a BH with a pericenter distance $r_{\rm p}>r_{\rm T}$ (where $r_{\rm T}$ is the tidal radius) are not tidally disrupted. 
However, they may still be tidally captured on more compact orbits around the BH as long as the total tidal energy dissipated during the first pericenter passage ($\Delta E_{\rm tid}$) exceeds the orbital energy at infinity ($T_\infty$).
In this appendix, we estimate the critical radius $r_{\rm C}$ for tidal capture.
In general, this parameter depends on the relative velocity between the BH and the star at infinity ($v_\infty$), as well as on the structure of the star (here modeled as a polytrope with polytropic index $n$).
Since captured stars eventually reach $r_{\rm T}$ through tidal energy dissipation at each pericenter passage and the cross section for capture is larger than the TDE cross section, the loss cone radius is effectively $r_{\rm C}$.

\begin{figure}
    \centering
    \includegraphics[width=0.5\textwidth]{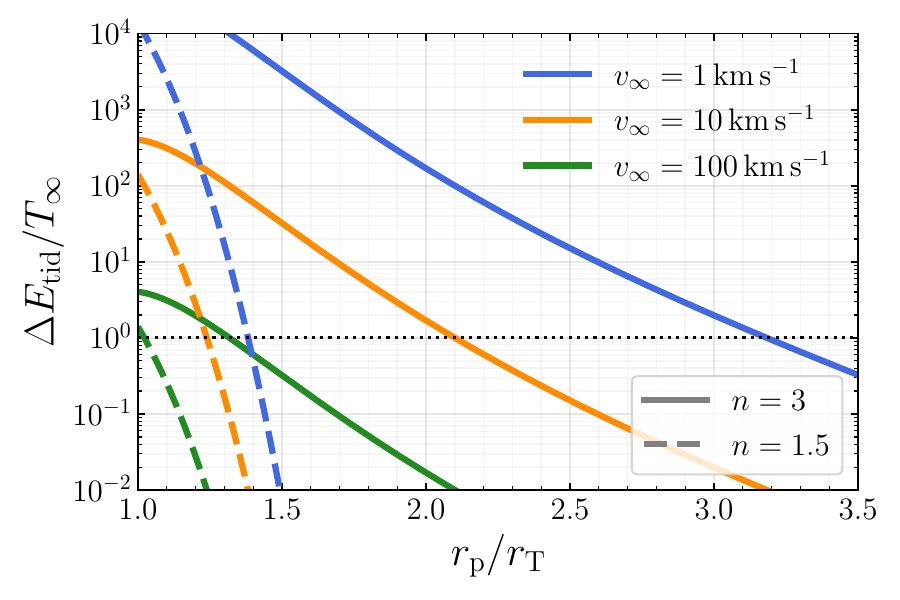}
    \caption{Tidal energy dissipated during the first pericenter passage $\Delta E_{\rm tid}$, including the quadrupole and octupole contributions, normalized to the energy at infinity ($T_\infty$). We show $\Delta E_{\rm tid}/T_\infty$ as a function of the pericenter distance relative to the tidal radius for two values of the polytropic index, $n=3$ (solid) and $n=1.5$ (dashed), and three values of the relative velocity at infinity. The horizontal dotted line corresponds to the threshold for tidal capture ($\Delta E_{\rm tid}/T_\infty=1$).}
    \label{Fig:dEt_rp}
\end{figure}

In Fig.~\ref{Fig:dEt_rp} we compute $\Delta E_{\rm tid}$ and plot it as a function of $r_{\rm p}/r_{\rm T}$ using the semianalytic fits of Ref.~\cite{1993A&A...280..174P}, including the multipoles $\ell=2$ and $\ell=3$ (quadrupole and octupole), for two polytropic indices: $n=3$ and $n=1.5$.
The point where $\Delta_{\rm tid}/T_{\infty}$ crosses unity corresponds to the critical pericenter for tidal capture. For pericenter distances larger than $r_{\rm C}$ the interaction is simply a flyby with the star escaping back to infinity.
As we have normalized the pericenter by $r_{\rm T}$ on the horizontal axis of Fig.~\ref{Fig:dEt_rp}, our computed ratio $\Delta_{\rm tid}/T_{\infty}$ has an extremely weak dependence on the BH mass, which can safely be neglected. Hence, we show results for $M_{\rm BH}=100M_\odot$ as a representative case.

The three-dimensional velocity dispersion of the systems that produce IMBHs with $M_{\rm BH}$ in the hundreds of $M_\odot$ varies in the range $\sim1$--$100\,\rm km\, s^{-1}$, with a typical value of $\simeq10\,\rm km\, s^{-1}$.
Assuming that the IMBH sits at the center of the cluster, 
$v_\infty$ can be well approximated by the stellar three-dimensional velocity dispersion. 
This is a reasonable assumption, because the Brownian motion of a central BH with mass $M_{\rm BH}$ has a wandering-to-core radius of $\simeq\sqrt{8m_\star/(3\pi M_{\rm BH})}$~\cite{1980ApJ...242..789L}, and thus the position does not fluctuate much in the case of IMBHs.
Our main focus is on tidal captures of main sequence solar stars, which are better described by a polytrope with $n=3$. From the solid orange line in Fig.~\ref{Fig:dEt_rp} we conclude that the critical pericenter distance for capture is typically at $\sim2.0r_{\rm T}$.
Therefore in the main text we use $r_{\rm C}=2.0r_{\rm T}$ when computing the loss cone radius.

\begin{figure}
    \centering
    \includegraphics[width=0.49\textwidth]{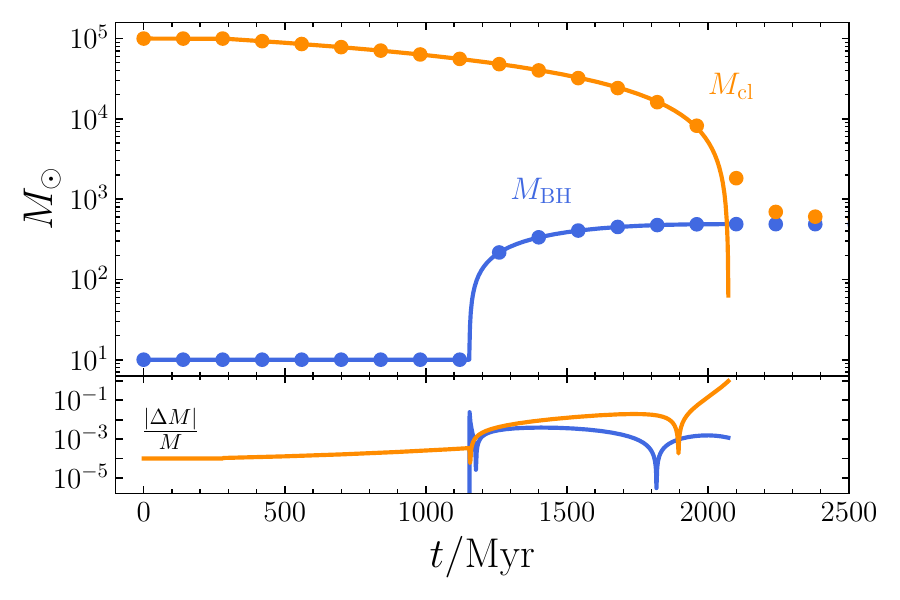}
    \caption{Top panel: the data points show the total cluster mass $M_{\rm cl}$ (in orange) and the BH mass $M_{\rm BH}$ (in blue) found by numerically solving the differential system of Eq.~\eqref{Eq:complete_differential_system} with initial conditions $M_{\rm cl,0}=10^5M_\odot$, $r_{\rm h,0}=1.2$, and assuming tidally limited evolution. The solid lines show the analytical approximation. For the purposes of this comparison, stellar evolution has been neglected. Bottom panel: relative error between the numerical and analytical solutions.}
    \label{Fig:numerical_analytical_comparison}
\end{figure}

\section{Cluster dissolution and BH feedback}
\label{App:Cluster_dissolution_and_BH_feedback}

In the isolated H\'enon's model, the ejection rate is low (about 0.74\% of stars are ejected every relaxation time~\cite{1985IAUS..113..521A}), and the systems of interest do not evaporate within a Hubble time. Moreover, we have verified that the BH-to-cluster mass ratio in this scenario always remains below $4\%$ and typically at the subpercent level, and the number of stars consumed by the BH is a tiny fraction of the total number. 
Thus the evolution of the global system approximately decouples from that of the BH, and the equations for $(dN_{\star}/dt, dr_{\rm h}/dt)$ and $d{M}_{\rm BH}/dt$ separate. 
This separation allows us to find a closed-form solution for the evolution of the system (notice that $dN_\star/dt$ and $dr_{\rm h}/dt$ are still coupled through the relaxation time). 
The growth of the BH strongly depends on the evolution of the cluster as a whole through the evolution of the velocity dispersion: see Eq.~\eqref{Eq:consumption_rate_b}. 

In the tidally limited model, however, the ejection rate is high enough for a cluster with typical initial conditions (say, $M_{\rm cl,0}\sim10^5M_\odot$ and $r_{\rm h,0}\sim1\,\rm pc$) to evaporate well within $\tau_{\rm Hub}$.
Therefore, at some point, the mass in stars reduces to the point where it is comparable to $M_{\rm BH}$, and the approximation $M_{\rm cl}\simeq m_\star N_\star$ fails.
In this appendix we discuss the applicability of this approximation (used in Sec.~\ref{Sec:Star_cluster_evolution}) to the tidally limited case.

The feedback of the BH into the evolution of the global properties of the cluster enters through the contribution of the consumption rate in removing stars from the system, as well as the contribution of $M_{\rm BH}$ to $M_{\rm cl}$. 
In general, the full set of equations for the evolution of the system becomes
\begin{subequations}
    \begin{align}
    M_{\rm cl} &= m_\star N_\star + M_{\rm BH}\,, \\
    {dN_\star\over dt} &= -\xi_{\rm e}{N_\star\over\tau_{\rm rh}} - \Gamma_{\rm C}\,, \label{eq:B2} \\
    {dr_{\rm h}\over dt} &= \zeta {r_{\rm h}\over \tau_{\rm rh}} + 2{r_{\rm h}\over M_{\rm cl}}{dM_{\rm cl}\over dt}\,, \\
    {dM_{\rm BH}\over dt} &= f_{\rm s}m_\star \Gamma_{\rm C}\,,
    \end{align}%
    \label{Eq:complete_differential_system}%
\end{subequations}%
where the consumption rate $\Gamma_{\rm C}$ is given by Eq.~\eqref{Eq:consumption_rate}, $m_\star=1M_\odot$ and $f_{\rm s}=0.50$ are assumed to be constants, and the other variables have been introduced in Sec.~\ref{Sec:Star_cluster_evolution}.
In Sec.~\ref{Sec:Star_cluster_evolution} we found an analytical solution to this differential system by dropping the second term. This assumption is justified when $M_{\rm cl}/M_{\rm BH}\gg 1$.

\begin{figure*}
    \centering
    \includegraphics[width=0.49\textwidth]{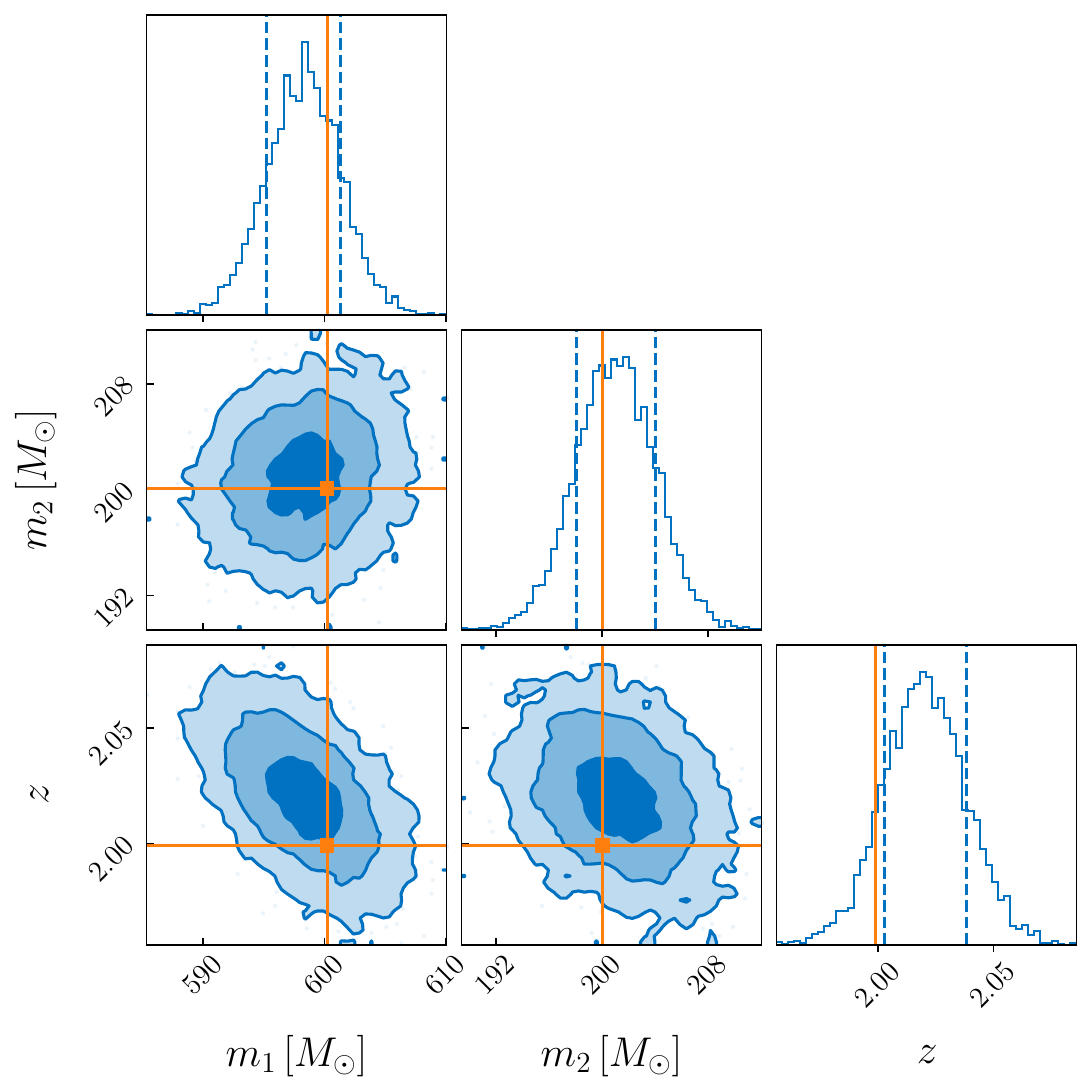}
    \includegraphics[width=0.49\textwidth]{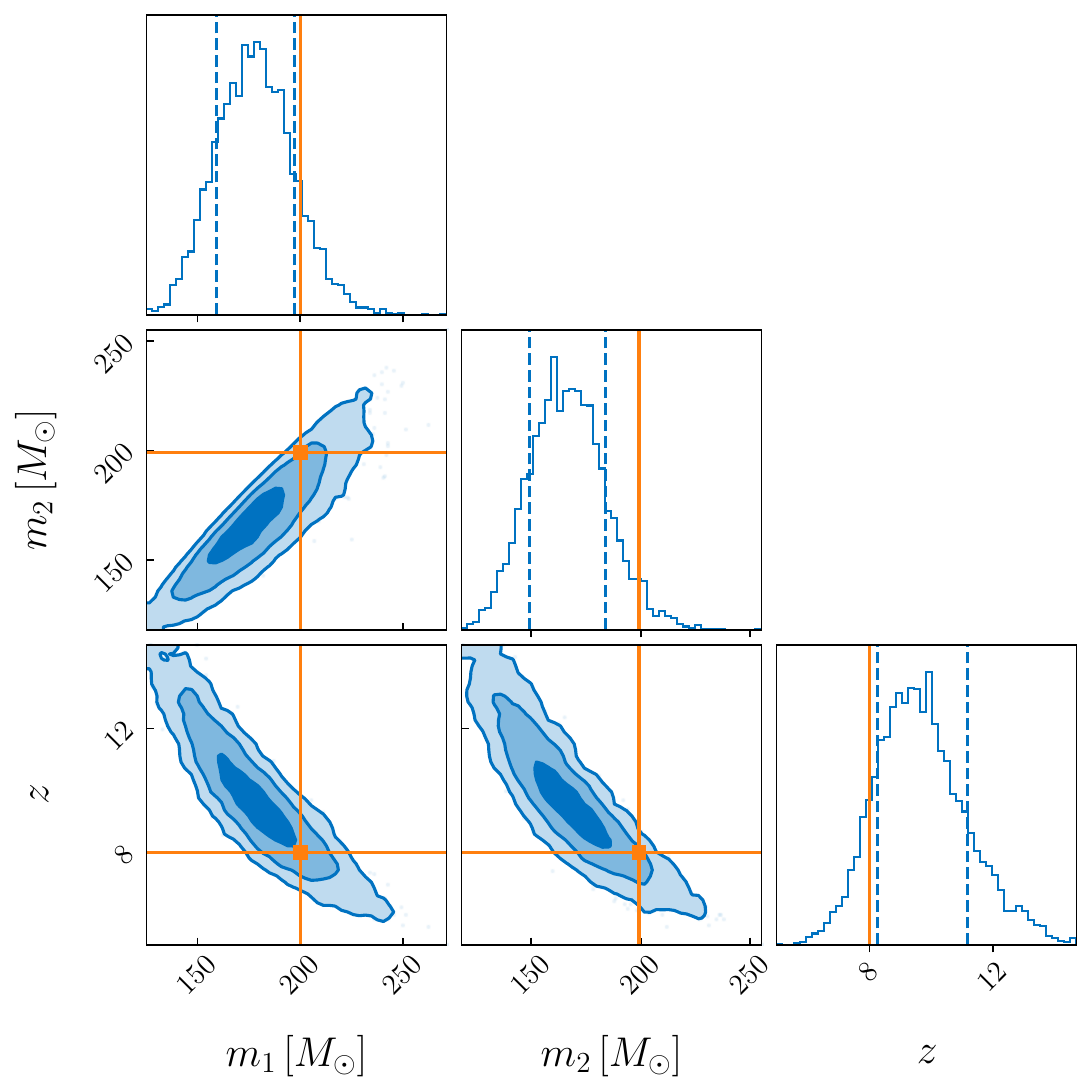}
    \caption{Posterior distributions of the source-frame masses and redshift for two representative binaries (cases D and F in Table~\ref{tab:binary_parameters}, respectively). The left panels show results for a binary BH with the following injected parameters:  $m_1=600\,M_\odot$, $m_2=200\,M_\odot$, $z=2$, $\chi_1=0.1$, $\chi_2=0.1$, $\theta_1=-0.65$, $\theta_2=1.06$, $\phi_{12}=2.81$, $\phi_{\rm JL}=2.26$, $\theta_{\rm JN}=2.22$, $\alpha=5.88$, $\delta=0.11$, $\psi=2.26$.  The right panels show results for a binary BH with parameters  $m_1=200\,M_\odot$, $m_2=200\,M_\odot$, $z=8$, $\chi_1=0.1$, $\chi_2=0.1$, $\theta_1=0.28$, $\theta_2=1.00$, $\phi_{12}=1.23$, $\phi_{\rm JL}=0.97$, $\theta_{\rm JN}=2.05$, $\alpha=5.28$, $\delta=0.65$, $\psi=0.97$. All angles are given in radians.}
    \label{fig:full_pe}
\end{figure*}

Here we numerically solve the system~\eqref{Eq:complete_differential_system} using a fourth-order Runge-Kutta method with initial conditions $N_{\star,0}=10^5$, $r_{\rm h,0}=1\,\rm pc$, and $M_{\rm BH,0}=10M_\odot$ in the tidally limited scenario, where $\xi_{\rm e}=0.045$ and $\zeta=0.0725$.
The top panel of Fig.~\ref{Fig:numerical_analytical_comparison} shows the time evolution of $M_{\rm cl}$ (orange) and $M_{\rm BH}$ (blue). The data points are found from numerical integration with time step $dt=10^{-2}\,\rm Myr$, which we found to ensure convergence with a relative error of no more than 0.1\%; the solid lines correspond to the analytical approximation.
The two solutions for $M_{\rm cl}$ diverge when the cluster evaporates after $t\simeq2000\,\rm Myr$, and the contribution of the BH mass to $M_{\rm cl}$ cannot be ignored in that regime.
However, the mass of the BH approaches its asymptotic value well before it becomes comparable to $M_{\rm cl}$. Therefore, the BH mass $M_{\rm BH}$ is well predicted by the approximate analytical model. 
The bottom panel of Fig.~\ref{Fig:numerical_analytical_comparison} shows the relative error in cluster and BH mass between the approximate analytical model and the full numerical solutions, and it illustrates that $M_{\rm BH}$ is predicted by the approximate model to an accuracy of $|\Delta M_{\rm BH}|/M_{\rm BH}\sim10^{-3}$. 

The second term on the right-hand side of Eq.~\eqref{eq:B2} corresponds to the reduction in the number of stars due to the BH. 
This term is important only in the initial phases of BH growth, dominating over ejections driven by relaxations for a short period of time (i.e., not many relaxation times).
In fact, we find that the cumulative relative change in $N_\star$ due to 
the consumption rate term from the runaway growth phase of the BH is only $\simeq0.5\%$. 
On the other hand, the first term of Eq.~\eqref{eq:B2} is important throughout the entire evolution of the cluster, affecting $N_\star$ over an extended period. 
The effect of this term becomes evident after several initial relaxation times, ultimately leading to cluster evaporation at around $t\simeq2100\,\rm Myr$.

\section{Parameter estimation of IMBH binaries}
\label{App:Parameter_estimation_of_IMBH-IMBH_binaries}

Here we describe the details of the individual Bayesian PE runs for the IMBH binaries considered in our study.

We choose a network of three next-generation GW detectors, consisting of one CE detector with 40\,Km arm length, one CE detector with 20\,Km arm length, and ET in the triangular configuration~\cite{Branchesi:2023mws}. The sensitivity, location, and orientation of the two CE detectors correspond to the CE-A and CE-B configurations of Ref.~\cite{Gupta:2023lga}, respectively. We assume ET to be at the current location of Virgo in Italy~\cite{VIRGO:2014yos}. We set the low-frequency sensitivity cutoff at $3\,\rm{Hz}$ for all the detectors in our network. 

We generate the GW signals using the \texttt{IMRPhenomXPHM} waveform model~\cite{Pratten:2020ceb}, a state-of-the-art quasicircular model that includes both spin precession effects and higher-order modes. Higher-order modes are expected to be important for both detection~\cite{Chandra:2020ccy} and PE~\cite{Fairhurst:2023beb} of systems with IMBH components. For each IMBH binary, we perform Bayesian PE on the whole set of 15 parameters that characterize the waveform model, namely,
\begin{eqnarray}
&&\{ 
M_z, 
q, 
D_L, 
\chi_1, 
\chi_2,
\theta_1, 
\theta_2, \nonumber\\ 
&&\phi_{12}, 
\phi_{\rm JL},
\theta_{\rm JN},
\alpha,
\delta,
\psi,
\phi_c,
t_c
\, \} \,.
\label{eq:params}
\end{eqnarray}
Here, $M_z=(1+z)M$ is the detector-frame total mass, $q=m_1/m_2 \geq 1$ the mass ratio, $D_L$ the luminosity distance, $\chi_{1,2}$ the spin magnitudes, $\theta_{1,2}$ the angles between the spins and the orbital angular momentum, $\phi_{12}$ the angle between the in-plane spin components, $\phi_{\rm JL}$ the angle between the total and the orbital angular momentum, $\theta_{\rm JN}$ the orientation angle, $(\alpha,\delta)$ the right ascension and declination, $\psi$ the polarization angle, $(\phi_c, t_c)$ the coalescence phase and coalescence time. We perform the PE runs using the public {\sc python} package {\sc BILBY}~\cite{Ashton:2018jfp,Krishna:2023bug}. We then convert the resulting posterior samples from the set of parameters $(M_z, q, D_L)$ to $(m_1, m_2, z_{\rm m})$, where $m_{\rm 1,2}$ denotes the source-frame IMBH masses.

We fix the spin magnitudes to be $\chi_{1,2}=0.1$ for all binaries. This value is consistent with the expectations for IMBHs formed through repeated TDEs (see Appendix~\ref{App:mass_spin_evolution}). Furthermore, in this study, we are mainly interested in constraints on masses and redshifts, which are not expected to be drastically affected by spins~\cite{Biscoveanu:2021nvg}. For each binary, we sample all the angular parameters isotropically at a reference frequency of $3\,\rm{Hz}$. 

In Fig.~\ref{fig:full_pe} we show PE results for two representative binaries in our study, i.e., cases D and F in Table~\ref{tab:binary_parameters}. 
We only show the posterior distributions for the main parameters of interest, i.e., the source-frame component masses and the redshift.

\begin{figure}
    \centering
    \includegraphics[width=0.49\textwidth]{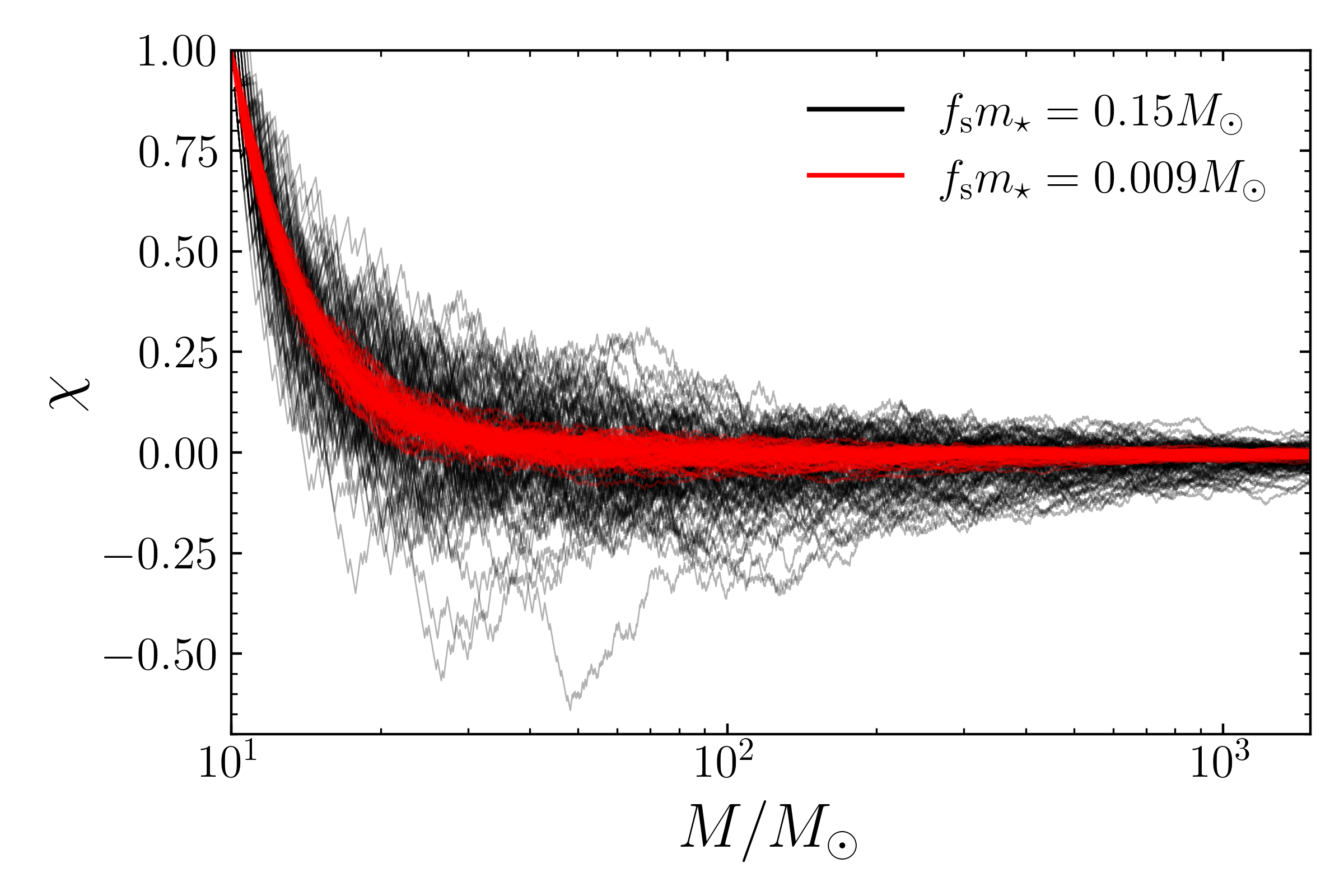}
    \caption{Stochastic evolution of the BH spin magnitude as it grows through the repeated consumption of stars using the approximation of Eq.~\eqref{Eq:update_spin_approximation}. The initial BH spin is always assumed to be maximal, while the mass consumed at each TDE is either $f_{\rm s}m_\star=0.15M_\odot$ (black lines) or $f_{\rm s}m_\star=0.009M_\odot$ (red lines). We simulate 100 realizations in each case.}
    \label{Fig:mass_spin_evolution}
\end{figure}

\begin{figure*}
    \centering
    \includegraphics[width=\textwidth]{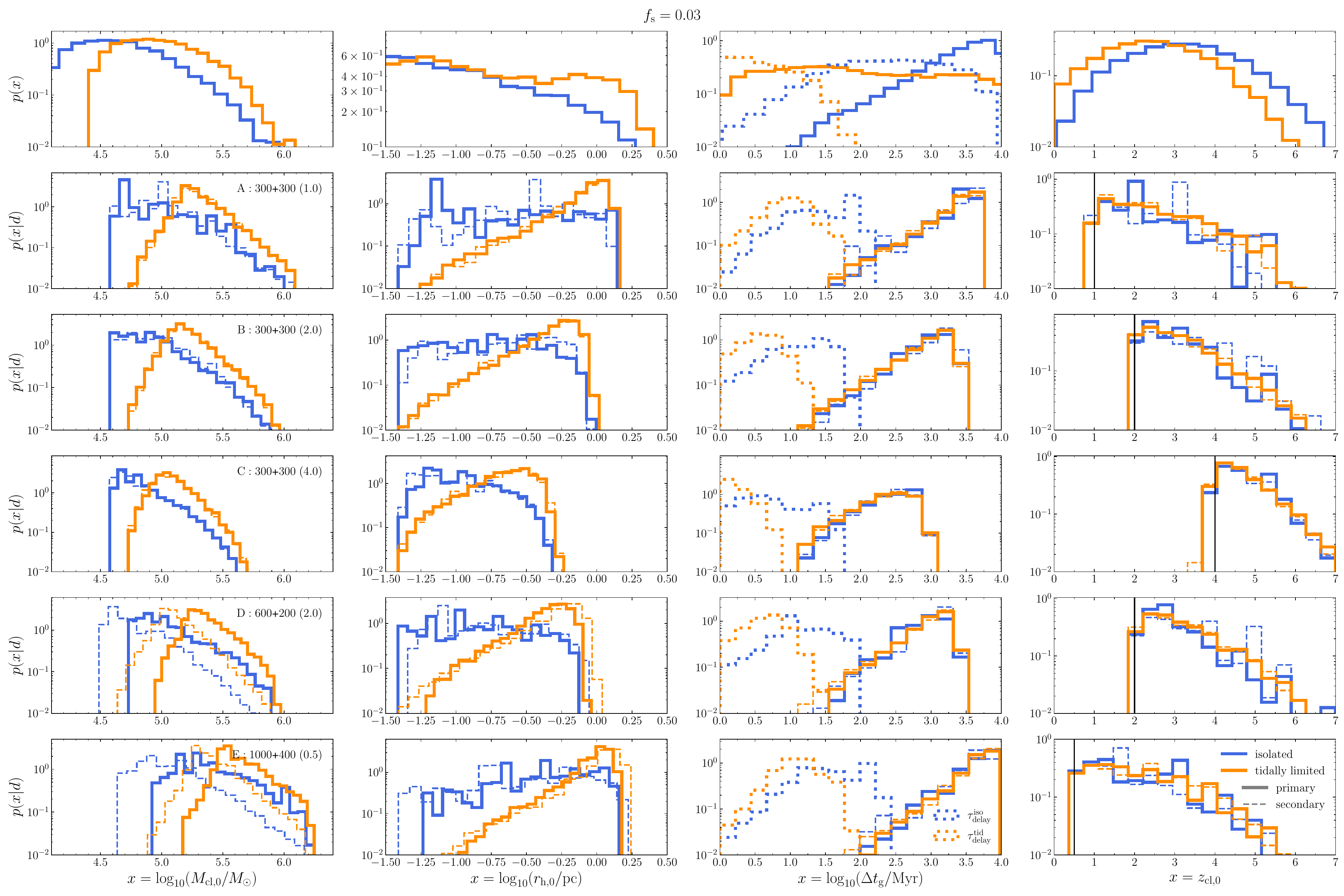}
    \caption{Same as Fig.~\ref{Fig:cluster_posteriors_lowz_fs05}, but with a lower accretion rate $f_{\rm s}=0.03$.}
    \label{Fig:cluster_posteriors_lowz_fs003}
\end{figure*}

\begin{figure*}
    \centering
    \includegraphics[width=\textwidth]{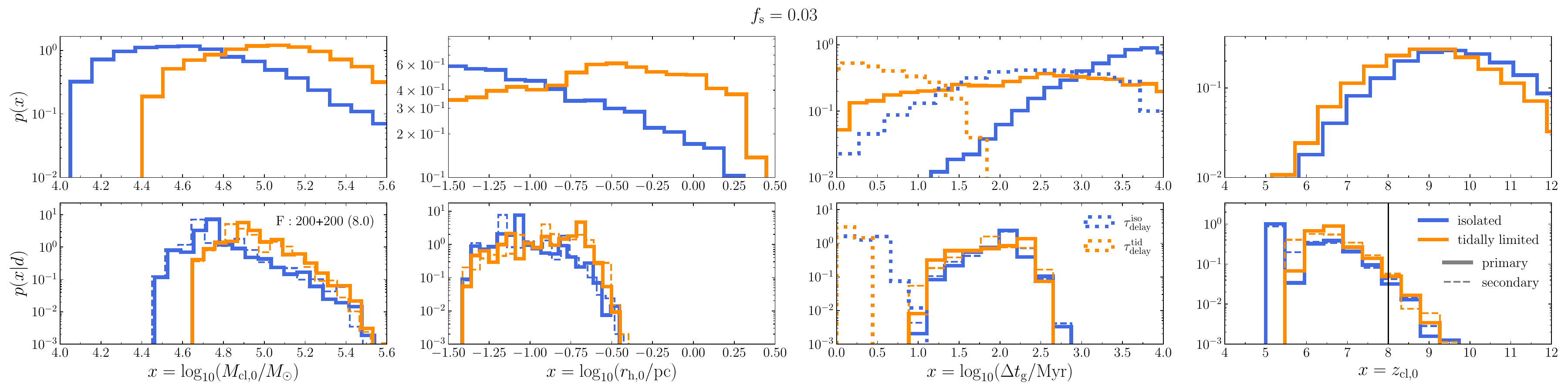}
    \caption{Same as Fig.~\ref{Fig:cluster_posterior_highz_fs05}, but with a lower accretion rate $f_{\rm s}=0.03$.}
    \label{Fig:cluster_posteriors_highz_fs003}
\end{figure*}

\section{Black hole mass-spin evolution}
\label{App:mass_spin_evolution}

Here we demonstrate that the spin of a BH growing through runaway stellar consumption asymptotes to zero. 

Consider a BH of mass $M$ and angular momentum $J$. 
The energy $\cal E$ and angular momentum $\Phi$ of a unit mass orbiting the BH at the innermost stable circular orbit (ISCO) are given by Eqs.~(2) and~(3) of Ref.~\cite{1970Natur.226...64B}, respectively. A mass element $\delta$ is accreted onto the BH once it crosses the ISCO radius. Then the mass of the BH increases by $\Delta M={\cal E}\times\delta$ while the angular momentum changes by $\Delta J=s \Phi \times\delta$, where $s=1$ ($s=-1$) for prograde (retrograde) accretion, corresponding to an increase (decrease) in $J$, respectively. 
Therefore we have $\Delta J / \Delta M = s \Phi / {\cal E}=s(GM/c^3){\cal G}(\chi, s)$, where
\begin{align}
    {\cal G}(\chi, s)\equiv{2\over3\sqrt{3}}{1 + 2\sqrt{3{\cal Z}(\chi, s) - 2}\over \left[1 - {2\over3{\cal Z}(\chi, s)}\right]^{1/2}}\,.
\end{align}
Here ${\cal Z}(\chi, s)$ denotes the ISCO radius normalized to the gravitational radius of the BH, $GM/c^2$.
It can be shown by differentiation that the equation for the evolution of the spin parameter, defined by $\chi\equiv cJ/(GM^2)$, is
\begin{align}
    {d\chi\over dM} = {s\over M}{\cal G}(\chi, s) - {2\chi\over M}\,. 
\end{align}
Integrating by separation of variables, we find
\begin{align}
    \int_{\chi}^{\chi'} {d\tilde{\chi} \over s{\cal G}(\tilde{\chi}, s) - 2\tilde{\chi}} = \ln{M'\over M}\,.
\end{align}
Assuming the BH has consumed a fraction $f_{\rm s}$ of a star of mass $m_\star$ after a TDE such that the accreted mass is $f_{\rm s}m_\star\ll M$, we can make the approximation that the spin varies only by a small amount. 
In the late phases of an evolving star cluster, the mean stellar mass is around $0.3M_\odot$ and $f_{\rm s}=0.5$ or $0.03$, such that $f_{\rm s}m_\star=0.15M_\odot$ or $0.009M_\odot$, respectively. 
In general, the BH needs to accrete an amount of matter of the order of its initial mass for the spin to change significantly~\cite{1970Natur.226...64B}. 
Under this approximation, $\ln (M'/M)\simeq f_{\rm s}m_\star / M$ to lowest order (which becomes much less than 0.01 as the BH grows beyond $10M_\odot$), and thus we derive a simple expression for the spin $\chi'$ of the BH after a single TDE:
\begin{align}
    \chi' \simeq \left( 1 - 2{f_{\rm s}m_\star\over M} \right)\chi + s {f_{\rm s}\over M}{\cal G}(\chi, s),
    \label{Eq:update_spin_approximation}
\end{align}
while the mass after the single TDE event is $M'=M + f_{\rm s}m_\star$. 
In Fig.~\ref{Fig:mass_spin_evolution} we show the evolution of the BH spin as a function of its mass as it grows through runaway tidal encounters with stars, assuming the ``worst case scenario'' in which the initial BH is maximally spinning. 

While formally the assumption of isotropy is broken by the spin of the BH, the BH spin effect is subdominant in our case, because the tidal radius $r_{\rm T}$ of a main-sequence star is many orders of magnitude larger than the gravitational radius, and the effect of the spin enters as an $O((GMc^{-2}/r_{\rm T})^3)$ term in the force (see e.g.~\cite{Hartle:2003yu}).
Therefore we assume prograde and retrograde accretion to be equally probable, as appropriate for a (nearly) spherically symmetric system.
The BH spin parameter $\chi$ asymptotes to zero as a consequence of the asymmetry in the ISCO radius between prograde and retrograde orbits:
retrograde orbits deliver a larger amount of (negative) angular momentum to the BH than prograde orbits. 
A similar conclusion is reached if the BH grows through repeated mergers with smaller BHs~\cite{Hughes:2002ei,Berti:2008af}. 
Qualitatively, $\chi$ decreases as the BH mass increases. 
The scatter in the spin evolution depends on the amount of mass consumed at each TDE. 
Starting with stellar-mass BH seeds that grow through the repeated consumption of low-mass stars, the spin magnitude of the BH becomes less than $\sim0.3$ by the time it grows beyond $100M_\odot$, and then the BH spin can safely be ignored. 
Incidentally, these considerations can be used to infer the growth scenario of BHs in their last mass-doubling epoch from BH spin measurements~\cite{Reynolds:2020jwt}. 

\section{Marginalized cluster posteriors for $f_{\rm s}=0.03$}
\label{App:Margninalized_cluster_posteriors_for_fs003}

In Figs.~\ref{Fig:cluster_posteriors_lowz_fs003} and~\ref{Fig:cluster_posteriors_highz_fs003}, for completeness, we plot the marginalized cluster posteriors for all of the IMBH binary systems listed in Table~\ref{tab:binary_parameters}.

These figures are similar to Figs.~\ref{Fig:cluster_posteriors_lowz_fs05} and~\ref{Fig:cluster_posterior_highz_fs05}, but with a lower fraction of each star's mass consumed by the BH ($f_{\rm s}=0.03$, rather than $f_{\rm s}=0.5$). 

\clearpage

\bibliography{inferring_cluster_from_IMBH}

\end{document}